\documentclass[a4paper,cleveref,autoref,thm-restate]{lipics-v2021}

\usepackage[dvipsnames]{xcolor}
\usepackage[framemethod=TikZ]{mdframed}
\usepackage{wrapfig}
\usepackage{lstcustom}
\usepackage{cite}
\usepackage{xspace}
\usepackage{enumerate}
\usepackage{cclicenses}
\usepackage{pifont}% http://ctan.org/pkg/pifont
\usepackage{booktabs}
\usepackage{todonotes}
\usepackage{url}
\usepackage{lineno,hyperref}
 \usepackage{siunitx}

\definecolor{firebrick}{rgb}{0.7, 0.13, 0.13}
\definecolor{mountainmeadow}{rgb}{0.19, 0.73, 0.56}

\definecolor{lgray}{gray}{0.9}
\definecolor{carmine}{rgb}{0.59,0.0,0.09}

\newmdenv [ %
 align=center, 
 skipabove=\topsep,
 skipbelow=\topsep,
 leftmargin       = 0.2              ,
 rightmargin      = 0.2              ,
 splittopskip     = \topskip      ]{mh}

\newcommand{\cmark}{{\color{mountainmeadow}\ding{51}}}%
\newcommand{\xmark}{{\color{firebrick}\ding{55}}}%

\newcommand{\mcsl}{\textsc{Meta-CrySL}\xspace}
\newcommand{\csl}{\textsc{CrySL}\xspace} 
\newcommand{\ca}{\textsc{CogniCrypt}$_{\textit{\textsc{sast}}}$\xspace}
\newcommand{\rascal}{Rascal-MPL\xspace}

\newcommand{\apidiff}{\textsc{apidiff}\xspace}

\definecolor{light-gray}{gray}{0.85}

\newcommand{\vt}[1]{\colorbox{light-gray}{\emph{#1}}}

\newcommand{\vbsc}{\vt{Variability on the Base Specification Class}\xspace}
\newcommand{\vsc}{\vt{Variability on Set Constraints}\xspace}
\newcommand{\vces}{\vt{Variability on Event Sets}\xspace}
\newcommand{\vmsr}{\vt{Modular Selection of \csl Rule}\xspace}

  \newcommand{\rqcode}{How many lines of \csl code can one save when writing
  \mcsl specifications?}

  \newcommand{\rqdup}{How much duplication of specifications is eliminated by using \mcsl in comparison to \csl?}

  \newcommand{\rqviolations}{What are the implications of instantiating
  \csl rules from \mcsl specifications, observing the
  number of API misuses \ca analysis reports?}

\lstdefinelanguage{CrySL}[]{Java}{
  morekeywords={ABSTRACT, SPEC, OBJECTS, EVENTS, ORDER, CONSTRAINTS, REQUIRES, ENSURES, REFINES, define, add, constraint},
  moredelim=[is][\textcolor{darkgray}]{\%\%}{\%\%},
  moredelim=[il][\textcolor{darkgray}]{§§}
}

\lstdefinelanguage{configuration}[]{Java}{
  morekeywords={config, src, out, load},
  moredelim=[is][\textcolor{darkgray}]{\%\%}{\%\%},
  moredelim=[il][\textcolor{darkgray}]{§§}
}

\newif\ifsplc

\title{Dealing with Variability in API Misuse Specification}

\author{Rodrigo Bonifácio}{Computer Science Department at University of Bras\'{i}lia}{rbonifacio@unb.br}{http://orcid.org/0000-0002-2380-2829}{funded by FAP-DF (research grant 05/2018)}

\author{Stefan Kr\"{u}ger}{Independent Researcher}{stefan.krueger@uni-paderborn.de}{}{}
\author{Krishna Narasimhan}{Technical University of Darmstadt}{kri.nara@st.informatik.tu-darmstadt.de}{}{}
\author{Eric Bodden}{Paderborn University \& Fraunhofer IEM}{eric.bodden@uni-paderborn.de}{http://orcid.org/0000-0003-3470-3647}{}
\author{Mira Mezini}{Technical University of Darmstadt}{mezini@informatik.tu-darmstadt.de}{}{}

\authorrunning{R. Bonif\'{a}cio, S. Kr\"{u}ger, K. Narasimhan, E. Bodden, and M. Mezini}



\funding{This research was supported by the DFG's collaborative research center 1119 CROSSING and ATHENE National Research Center
  for Applied Cybersecurity}

\Copyright{Rodrigo Bonif\'{a}cio, Stefan Kr\"{u}ger, Krishna Narasimhan, Eric Bodden, Mira Mezini}

\EventEditors{Manu Sridharan and Anders M{\o}ller}
\EventNoEds{2}
\EventLongTitle{35th European Conference on Object-Oriented Programming (ECOOP 2021)}
\EventShortTitle{ECOOP 2021}
\EventAcronym{ECOOP}
\EventYear{2021}
\EventDate{July 12--16, 2021}
\EventLocation{Aarhus, Denmark (Virtual Conference)}
\EventLogo{}
\SeriesVolume{194}
\ArticleNo{21}

\begin{document}

% ARXIV note: Before submitting to ARXIV, uncomment the
% following line.

\nolinenumbers

\maketitle
\begin{abstract}
APIs are the primary mechanism for developers to gain access to externally defined services and tools. However, previous research has revealed API misuses that violate the contract of APIs to be prevalent. Such misuses can have harmful consequences, especially in the context of cryptographic libraries. Various API-misuse detectors have been proposed to address this issue---including CogniCrypt, one of the most versatile of such detectors and that uses a language (CrySL) to specify cryptographic API usage contracts. Nonetheless, existing approaches to detect API misuse had not been designed for systematic reuse, ignoring the fact that different versions of a library, different versions of a platform, and different recommendations/guidelines might introduce variability in the correct usage of an API. Yet, little is known about how such variability impacts the specification of the correct API usage. This paper investigates this question by analyzing the impact of various sources of variability on widely used Java cryptographic libraries (including JCA/JCE, Bouncy Castle, and Google Tink). The results of our investigation show that sources of variability like new versions of the API and security standards significantly impact the specifications. We then use the insights gained from our investigation to motivate an extension to the CrySL language  (named MetaCrySL), which builds on meta-programming concepts. We evaluate MetaCrySL by specifying usage rules for a family of Android versions and illustrate that MetaCrySL can model all forms of variability we identified and drastically reduce the size of a family of specifications for the correct usage of cryptographic APIs. 
\end{abstract}

\section{Introduction}

Application Programming Interfaces (APIs) have become fundamental to increase developer productivity.
Nonetheless, prior research~\cite{acar:sp2017,fischer:sp2017,nadi:icse2016}
has indicated that developers often struggle with using APIs for various reasons,
including poor documentation, low-level abstraction, and lack of tool support.
One way to mitigate these issues are approaches to detecting API misuses~\cite{amman:tse2017, kruger:tse2019, egele:ccs-2013, yongcryptotracer,findbugs}.
%, which fall in one of two categories.
Most of these approaches are deny-listing approaches~\cite{amman:tse2017,yongcryptotracer,egele:ccs-2013,findbugs}---providing analyses that scan for \emph{incorrect} API uses. Deny-listing approaches suffer from false negatives and cannot be easily extended because they rely on hard-coded rules~\cite{kruger:tse2019}. To address these issues, CogniCrypt~\cite{kruger:tse2019} follows an allow-list approach and instead of hard-coding what the correct usages are, it takes a set of correct usage rules as a parameter---the latter are 
specified by API developers using the \csl specification language~\cite{kruger:tse2019}.
%\sk{2 and 26 are only in the super list, 19 on the other is not part of the super list at all.}
For instance, \csl has been used to model correct usage rules of Java Cryptographic APIs.  

%\sk{This argument mixes up two (fairly) orthogonal things. An Allow-list approach may also use a specification language. It just happens so that before \csl a) no white-listing approaches for Crypto APIs existed and b) the black-listing ones we looked at were relying on hard-coded rules. \csl just combines the best of both.}

%use DSLs that enable domain experts to easily express the rules whose violations result in a misuse, and as a result avoid false positives that result from not capturing the intention of the expert.
However,  the correct usage of an API is often subject to various sources of variability. %, which affect their syntax and semantics. 
They include (but are not limited to) evolving signatures and behavioral changes, e.g., due to different security standards in case of crypto APIs.\footnote{Cryptographic libraries have different definitions of correctness----and in particular \emph{secure}---usages, based on the standards like FIPS or BSI under which they operate contributing to yet another source of variability.} Last but not least, APIs like Java Cryptography Architecture (JCA) 
%which is the standard library for performing cryptographic tasks in Java, 
foster flexibility 
%over other constraints (such as API usability). Flexibility is promoted 
through the use of different \emph{providers} that can be \emph{plugged 
into} to override the default implementation of an algorithm. 
%However, this degree of flexibility 
%to use different algorithms 
%can pose difficulties to maintain the specifications in API usage pattern languages 
%like \csl. 
Depending on the JCA provider, different secure algorithms (according to
cryptographic standards) might be available or not. 
Whether or not an API usage is correct may also vary owing to other factors, including version of the platform (e.g., Java Platform, Android Platform) and version of the API
implementations.
%, and the current standards that recommend a set of algorithms for use in applications. {\bf MM: standards have been mentioned above. I believe that overall the discussion of the sources of variability need some consolidation and streamlining...}
%The question how the variability in API usages has on allow-listing approaches like \csl has not been investigated so far. 
%
Nonetheless, there is a lack of understanding about 
%A systematic investigation of
(a) how sources of API variability affect what should be considered the correct 
usage of an API and (b) a solution to modelling this variability in allow-listing approaches like \csl. % are missing today.
This is where this paper makes its contributions.

%Specifically, we investigate the effects of variability for the Java cryptographic APIs and the \csl  specification language.
% for specifying correct API usage patterns as the subject of our investigation. 
We perform an in-depth domain engineering on the
correct usage of cryptographic APIs. 
To this end, we consider the following sources from which variability might originate from: cryptographic standards (FIPS, ECrypt, and BSI), cryptographic libraries (e.g., JCA, Google Tink, Bouncy Castle), cryptographic library implementations (e.g., JCA providers), and cryptographic library evolution. Based on the findings, we implement \mcsl, a meta-programming approach for managing families of \csl specifications, ensuring that different sources of variability can be accounted for when specifying usage patterns. 
Using the new set of specifications, we conduct
an empirical study to investigate two characteristics of
\mcsl: \emph{expressiveness} (i.e., \emph{the possibility to express all
sources of \csl variability using \mcsl}), \emph{compactness}
(i.e., \emph{number of lines of \csl code one can save when writing
\mcsl specifications} and the fraction of redundancy one can eliminate) and \emph{correctness} (i.e., \emph{does the specifications generated by \mcsl detect distinct violations when exploring different configurations of \csl rules}).
% These characteristics have been
% explored in a previous study to analyze domain-specific
% languages~\cite{Kahraman2015}.
%Commenting this as we also have the same feedback in evaluation section
%\eb{I would have expected some measure of reduction of duplication. Ideally each ``thing'' should be expressed only once.}

%In this work, we focus on Java crypto APIs and \csl.
%% for several reasons.
%% as subjects of our investigation:
%Java cryptographic APIs are subject to many sources of variability. 
%\csl is a comprehensive API specification language for the latter, which allows us to 
%compare our approach with \csl. \sk{That sounds like circular logic: "We use crypto apis and crysl because crysl was designed for crypt apis."}
%To the best of our knowledge, \csl is the state-of-the-art in API specification in terms of expressiveness and reducing duplication of specifications thereby reducing the effort of domain experts.
%
%Even though the scope of our investigation is specific to cryptographic APIs, 
We believe that one can also benefit from using a
domain engineering approach for specifying the correct
usage of non-cryptographic APIs as well. First because the sources of variability we
discuss are not unique to cryptographic APIs as all APIs offer variability in behavior due to
evolving signatures as a result of new versions. Second because variability as a result of pluggable
implementations from different providers is not unique to JCA,
either (c.f., JDBC~\footnote{\url{https://www.oracle.com/database/technologies/jdbc-migration.html}}).
Even security standards that are unique to cryptographic APIs have parallels in the form of context-specific usage patterns for non-crypto APIs. 

To summarize, the main contributions of this paper are as follows:
\begin{itemize}
\item Domain engineering on Java cryptographic libraries, including:
  \begin{itemize}
  \item A study on the evolution of Java cryptographic APIs.
  \item A study on different cryptographic standard recommendations.
  \item A discussion about how the evolution of cryptographic libraries and cryptographic recommendation impact on the
    correct usage of APIs.
  % \item A study on how open-source projects use Java cryptographic libraries.
  \end{itemize}   
  \item The design and implementation of \mcsl, an extension to \csl that helps manage sources of variability on \csl specifications. 
  \item An evaluation that shows how  \mcsl  can help API experts to better
    modularize variability in \csl specifications.   
\end{itemize}

In Section~\ref{sec:background}, we discuss some concepts that are pre-requisite to understanding the remainder of the paper. We present our analysis of sources of API variability in Section~\ref{sec:domainanalysis}. In Section~\ref{sec:design}, we present the design of \mcsl, the language that resulted out of the insights gained from our study. Lastly, we empirically evaluate \mcsl in Section~\ref{sec:eval}.

\section{Background} 
\label{sec:background}

In this section, we present the 
concepts and definitions necessary
to understand our research context, contributions,
and results. In Section~\ref{sec:crypto-apis}, we introduce the challenges
for using cryptographic APIs correctly. Although
we perform the first part of our study with different
cryptographic APIs, we will use the JCA to drive home these challenges in this section. In Section~\ref{subsec:cryptostds}, we present the cryptographic standards we consider in
our research. These standards may guide and impact the specifications of the
correct usage of cryptographic APIs. Finally, Section~\ref{sec:crysl} introduces
the \csl language,
which allows experts to specify the proper usage of Java cryptographic APIs.

\subsection{Cryptographic APIs}\label{sec:crypto-apis}

Ferguson et al.~\cite{ce-book} state that ``cryptography
is very difficult'', mostly because it involves several
branches of mathematics and computer science~\cite{ce-book,sl-book}.
For this reason, algorithms and implementations are only recommended after
a huge effort on testing---often conducted by a public
community. That is, regardless of how much they have been vetted,
they are at best \emph{still secure} or \emph{not yet insecure}.
As a result, developers should rely on
well-known cryptographic algorithms and API implementations that
are subject to hundreds or thousands of hours of cryptanalysis~\cite{sl-book}. 

Cryptographic APIs (or libraries) that exist for each major
programming language,such as JCA and Bouncy Castle for
Java and wolfCrypt and OpenSSL for C/C++, make
available a number of implementations for
performing cryptographic tasks, such as the support
for generating (pseudo) random numbers, message digests, symmetric and asymmetric cryptography (including digital signature).
Although these libraries share
similar characteristics, their design differ
according to distinct principles, such as flexibility and
usability. Unfortunately, existing
research reports that these APIs
are often complex and hard to
use~\cite{nadi:icse2016,acar:sp2017}, which in the end might
compromise the security of the systems.

For instance, JCA has been designed such
that it is possible to change the
cryptographic implementations used in a
system without having to modify many parts of
the system. 
Specifically, this API employs the provider architecture~\cite{JCARG} that enables implementations behind the interfaces to be easily swapped.
% \sk{I've changed the description of the provider model above because I considered the old one incorrect. If you disagree about this change, please message me.}
The official documentation of JCA~\cite{JCARG} explicitly mentions that the three main motivations driving the design of the API were:

\begin{enumerate}
	\item \textbf{Implementation independence: } Applications can choose between many variants of implementations of cryptographic algorithms
	\item \textbf{Implementation interoperability: } Just like the applications are not tied to providers, providers are also not tied to applications
	\item \textbf{Algorithm extensibility: } Cryptographic algorithms can use building block primitives from variable sources to compose their algorithms	
\end{enumerate}

Figure~\ref{fig:tc-md} shows a usage scenario for
the \texttt{MessageDigest} class of the
JCA, which computes a
hash of input data. The first step to this end is to
get an instance of an implementation using a string
that specifies the message digest algorithm (\texttt{BLAKE2B-512}),
and, optionally, a named reference to a provider that 
makes available the actual implementation of the algorithms through the
JCA interface. After getting a
\texttt{MessageDigest} instance, a developer might populate
the digest by calling the \texttt{update()} method one or more times, and then
calling the \texttt{digest} method to compute a hash value of the
input data. The same sequence of events has been valid since the first specification of this API.
However, several new message digest algorithms have been implemented (e.g., the family of SHA-3 algorithms has been introduced in Java 9).
Others have been deprecated and considered insecure over the years (e.g., algorithms MD2, MD5, and SHA-1 are
not recommended anymore~\cite{bsi}).

\begin{figure}[htb]
\begin{lstlisting}[language=Java]
@Test
public void testBlakeDigest() {
  try {
    MessageDigest md = MessageDigest.getInstance("BLAKE2B-512", "BC");
    md.update(data);
    byte[] res = md.digest();
    Assert.assertNotNull(res);
  }
  catch(Exception e) {
    org.junit.Assert.fail(e.getMessage());
  }
}
\end{lstlisting}
\caption{Code snippet for computing a message digest using the JCA
  Bouncy Castle provider (identified by the \texttt{BC} string)}
\label{fig:tc-md}
\end{figure}

Therefore, to correctly use JCA,
developers must not only understand the
expected sequence of method calls for
each cryptographic primitive, but which algorithms
and providers are available and are still considered secure.
Cryptographic standards detail which algorithms and
algorithm configurations developers should use while
implementing systems that deal with sensitive information. 
Given the complexity related to the use of 
crypto APIs, existing research uses static
analysis tools to assess the correct usage of crypto
APIs~\cite{kechagia:issta2019,kruger:tse2019,Rahman:ccs19} and code generation to
assist developers to correctly
implement cryptographic tasks~\cite{kruger:ase2017}. 

%From the above, it is clear why the cryptographic domain and especially the JCA is
%a rich scenario to explore techniques for accessing the impact of variability on the correct usage of APIs.

% \sk{Since we are discussing the standards relevant to our work explicitly below, should we also discuss the other libraries here, at least briefly?}

\subsection{Cryptographic standards}
\label{subsec:cryptostds}

A cryptographic standard details a set of recommendations related to the use of cryptographic primitives. A few examples of cryptographic standards include:

\begin{description}
\item[FIPS Standards] present a set of requirements from the American
  National Institute of Standards and Technology (NIST) that should be considered
  when implementing security modules for computational systems~\cite{fips}.
  This set of standards
  suggest algorithms for different primitives, including symmetric encryption, digital
  signatures, and message digest. 
  
\item[BSI TR-021-102-1] is a technical guideline from the German Federal Office for Information Security (BSI) that provides the results of a security assessment on cryptographic algorithms. This assessment supports a long-term orientation on the use of cryptographic mechanisms~\cite{bsi}.
  
\item[ECrypt TR-D5.4] details a set of recommendations about cryptographic algorithms and
  key size. It is an effort from the Ecrypt Coordination and Support Action, an initiative
  from the European Unions' H2020 program~\cite{ecrypt}. 
\end{description}

\subsection{\csl: Assessing the Correct Usage of Cryptographic APIs}\label{sec:crysl}

As can be seen from the above, applying cryptographic APIs within a software can have a lot of potential for errors and developers require new techniques and tools to support the use of cryptographic APIs. A research effort involving different institutions designed and developed CogniCrypt, a suite of tools that leverages the specification language \csl to enable API experts to specify the correct usage of libraries. \ca~\cite{kruger:tse2019} is a module of CogniCrypt that takes rules in \csl and a target program as input and uses state-of-the-art
data-flow
analysis~\cite{johannes:ecoop2016, johannes:oopsla2017,johannes:popl2019}   
to identify deviations from these rules in this program. 

In its current version,
\csl allows cryptographic experts to specify how
to instantiate and use an object-oriented class that
implements a cryptographic primitive. Figure~\ref{fig:csl-md-jca} shows the \csl specification
for the \texttt{MessageDigest} class of the JCA
API. 

\begin{figure}[htb]
\begin{lstlisting}[language=CrySL,escapechar=@]
SPEC java.security.MessageDigest
OBJECTS
  java.lang.String algorithm;
  byte[] data;
  byte[] digest;
EVENTS
    g1: getInstance(algorithm);
    g2: getInstance(digestAlg, _);
    
    Gets := g1 | g2;

    u1: update(_);
    d1: out = digest();
ORDER
    Gets, u1+, d1     
CONSTRAINTS
    algorithm in {"SHA-256", "SHA-384", "SHA-512", "BLAKE2B-512"};@\label{line:jca-alg}@

ENSURES
    digested[out];
\end{lstlisting}
\caption{\csl rule for the \texttt{MessageDigest} JCA API (considering the default provider)}
\label{fig:csl-md-jca}
\end{figure}

A \csl rule explicitly states the \emph{class under
specification} in the \texttt{SPEC} clause. The \texttt{OBJECTS}
definition describes a list of object declarations. These objects
might appear as arguments to events or as variables assigned to the 
return value of an event. The \texttt{EVENTS} section declares
the methods of the \emph{class under specification} that are
relevant for specifying the correct usage of the class.
In particular, the order in which these (labeled) methods should
be called appears as a regular expression in the \texttt{ORDER}
clause. Several operators can be used to denote this
regular expression. That is, supposing that we have events 
with labels $e_1$ and $e_2$, we can combine these events
using either the \emph{sequence operator} ($e_1, e_2$) or
the ``or'' operator ($e_1 \mid e_2$). We can
also state that one event is optional ($e_1$?) or that
an event might either occur zero or more times ($e_1\ast$) or one or
more times ($e_1$+). It is also possible to define
\emph{aggregates} (such as \texttt{Gets := g1 | g2}), which
help with the definition of the \texttt{ORDER} clause. The
example of Figure~\ref{fig:csl-md-jca} states that
a developer must first call one of the
\texttt{getInstance()} methods (using the \texttt{Gets} aggregate)
before calling the \texttt{update()} method at least once. After that,
the developer must conclude the computation of the message digest
using the \texttt{digest()} method. A \csl compiler
translates this regular expression into a state machine.
After that, the \ca~ component\cite{kruger:tse2019}  analyzes a system
to verify if a sequence of calls to a
\texttt{MessageDigest} instance obeys the expected
sequence of events of the \texttt{ORDER} clause.

The \texttt{CONSTRAINTS} clause allows a cryptographic
expert to define constraints on the objects declared
in a \csl rule. For instance, the \csl rule of
Figure~\ref{fig:csl-md-jca} states that the
\texttt{algorithm} used as parameter
for the \texttt{getInstance()} methods should be 
evaluated to one of the string literals that represent
a ``secure'' message digest algorithm supported
by the JCA default providers: \texttt{SHA-256}, \texttt{SHA-384},
or \texttt{SHA-512}. Therefore, during the analysis of
a system, \ca reports an error if it finds
a call to the \texttt{getInstance()} method of the
\texttt{MessageDigest} class using a different algorithm
(such as \texttt{MD5}). Finally, the \texttt{ENSURES} clause
of a \csl rule allows a cryptographic expert to state a
predicate that can be later used as a pre-condition in
a \csl specification for a different class (using the
\texttt{REQUIRES} construct of \csl).
There are other \csl constructs that we do
not discuss here, and a reader that is interested in
a more detailed description should read the paper that
introduced the \csl specification language~\cite{kruger:tse2019}. 

Previous studies have shown the efficiency of using
the \csl approach in identifying common misuses of
cryptographic APIs~\cite{kruger:tse2019}, \emph{but considering only one specific
set of \csl rules}. Nonetheless, as we discuss in
the remainder of this paper, \csl rules
should consider possible sources of variability
that might affect the specifications, including
versions of APIs and platforms and cryptographic
standards. 
 
\section{Domain Analysis}\label{sec:domainanalysis} 

To better understand the impacts of variability on API misuse specification, we conducted a \emph{domain analysis}~\cite{apel-spl-book,spl-book} that sought
to understand reuse opportunities across Crypto-API-usage specifications,
considering different libraries, their different providers and their different versions, different cryptographic primitives, and different
cryptographic standards---altogether corresponding to the \emph{sources
of variability} of our domain analysis.
Domain Analysis is a well-established set of activities
in the software product line community. The goal is to
identify variability motivating the implementation of an infrastructure for software
reuse~\cite{spl-book,apel-spl-book}. 

\subsection{Study Settings}\label{sec:settings}

We setup our study based on the following research questions:

\begin{enumerate}[RQ1]
\item \textbf{How do different APIs and their implementations
  (e.g., different JCA providers) vary
   the specifications
  of the correct usage of cryptographic primitives?}
  Motivation: Previous studies using specification languages like CrySL only considered
	the correct usage of the \emph{default} providers for the JCA.
	These studies report that almost 95\% of Android
	applications that use cryptographic APIs present at least one misuse of
	these APIs~\cite{kruger:tse2019}. Answering \textbf{RQ1} is relevant because alternative providers such as Bouncy Castle support algorithms that are not supported by the default providers. It is unclear whether findings of the previous \csl studies remain valid (particularly in the cases where an application explicitly uses a different provider). 
  
\item \textbf{How do existing cryptographic standards
  vary the notion of secure or compliant use of cryptographic libraries? }
  Motivation: Although the use of some cryptographic algorithms
        are considered insecure (e.g., \texttt{MD5}
	and \texttt{SHA-1}), they are still widely used in practice. There are many
	reasons for that, including compatibility with existing legacy code
	and the lack of knowledge of developers about up-to-date
	cryptographic algorithm recommendations. In addition, current security
	standards (such as FIPS and ECrypt) present recommendations about
	which algorithms should be used now and in a near future.
        Answering \textbf{RQ2} helps us to construct a baseline
        regarding how secure existing application are when considering existing
	standards. Moreover, \textbf{RQ2} helps to understand the relevance
	of security standards to the specification of the correct usage
	of cryptographic APIs.
 \item \textbf{How does the evolution of a cryptographic library
   vary its correct usage over time? }
   Motivation: Answers to \textbf{RQ3} will bring new insights about
	how to specify
	policies and static analysis tools that aim
	to guarantee the correct usage of cryptographic APIs,
	considering that they might evolve along the way. Moreover,
	answering \textbf{RQ3} might provide evidence that
	the evolution of APIs must be considered
	when specifying their correct usage.
\end{enumerate}

% \subsection{Research Methodology}\label{sec:method}

To answer the RQs,
we first conducted a domain analysis on the specification
of the correct usage of Cryptographic APIs.
We first read the documentation of APIs
and looked at code examples (including test cases)
that use Java (JCA, Bouncy Castle, and Google Tink) and
C/C++ (OpenSSL and wolfCrypt) cryptographic libraries.
We then built a general understanding about how different sources of
variability might influence our
domain, i.e., the domain of specification of the correct usage of
cryptographic APIs. 

\begin{figure*}[htb]
  \includegraphics[scale=0.35]{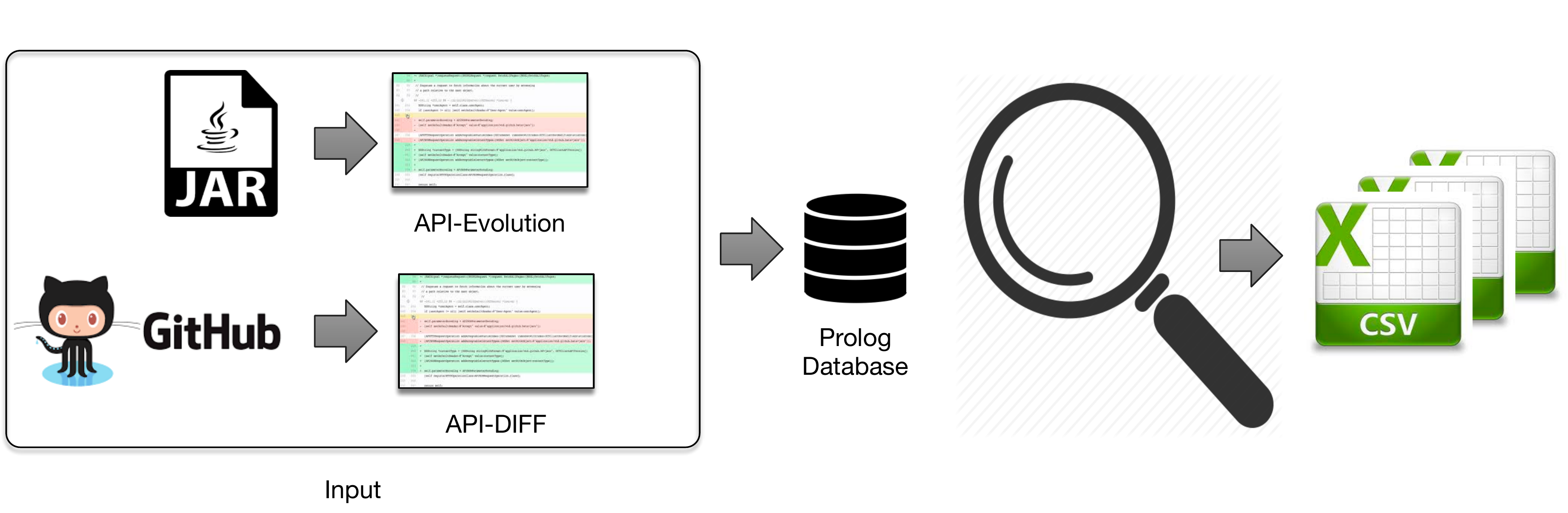}
  \caption{Approach for mining the evolution of Java Cryptographic APIs}
  \label{fig:api-evolution}
\end{figure*}

Figure~\ref{fig:api-evolution} shows the general workflow that we use
to mine the evolution of Java cryptographic libraries (JCA, Google Tink, and
Bouncy Castle). We leverage \apidiff~\cite{api-diff,brito:saner2018} and our own 
static analysis tool to mine classes and methods available per API release and the patterns
of changes along the evolution of the libraries. We populate all
this information into a Prolog database of facts and rules
that allow us to answer questions concerning both newly introduced algorithms as well as deprecated ones for specific versions of a given library. Introducing and removing new
primitive algorithms suggest that there should exist \csl rules for every version of that API that introduces a change. Using a customized version of \apidiff, we also
investigated breaking changes~\cite{brito:ese-2020,brito:saner2018,xavier:saner2017}, that is, changes between consecutive releases of an API that break the client code. Next, we execute
queries into this database and export the results to CSV files to analyze and
understand the evolution of crypto APIs.

%% We also mine the repository of open-source Java projects to understand
%% what cryptographic libraries are actually used in Java projects and
%% how developers use these libraries.   
%% Previous work~\cite{nadi:icse2016} has established that JCA is the most-widely
%% used API. To examine this claim, we mined 500 randomly selected Java and Android
%% projects on GitHub that use cryptographic APIs. For this study, we used the
%% same set of cryptographic libraries from
%% a previous research work~\cite{}. 
%% As around 97\% of them use the JCA, Bouncy Castle or both, for RQ2 and RQ3, we focus our discussion on these two APIs.}
%% \sk{While writing this, I noticed that we never introduce the Lightweight Bouncy Castle API and what is different about it compared to Bouncy Castle's JCA provider. We just start referring to it on page 9.

\subsection{Analysis Results}\label{sec:domain}
\paragraph*{RQ1: Variability Due to Different Cryptographic APIs}

We started our domain analysis by exploring
different cryptographic APIs (e.g., JCA, Google Tink, and
Bouncy Castle, Open SSL and wolfCrypt). We soon realized that these
APIs differ significantly in terms
of design principles and decisions. For instance, the design of the
JCA considers flexibility as a key element. Developers are responsible 
for specifying the configurations of keys and algorithms as well as modes of
operations they want to use, which has proven to be challenging to developers.
Certain configuration problems might only appear at runtime adding to the complication.
The design decisions of Google Tink, on the other hand, favor simplicity, instead of
flexibility. This way, there is a small set of 
key / algorithm configurations available, and the
developer is encouraged to use one of these configurations, in order to
avoid possible API misuses. 

In comparison to Google Tink, C/C++ libraries are yet more restricted. That is, 
OpenSSL and wolfCrypt define specific functions for
each algorithm. The code snippet in Figure~\ref{fig:md-wolf} shows
how to use the wolfCrypt library to generate a hash of an input data
using the Blake2b algorithm. There are several
calls to functions that are specific to this algorithm.
Since the different implementations of
message digest algorithms in wolfCrypt do not share a common
interface, the code of the \texttt{digestData} function is not flexible.
In the case a developer has to change the message digest algorithm,
she would have to rewrite the entire function.

\begin{figure}
\begin{lstlisting}[language=C]
byte[64] digestData(byte input[64]) {
  byte digest[64];
  Blake2b b2b;
  wc_InitBlake2b(&b2b, 64);
  wc_Blake2bUpdate(&b2b, input, sizeof(input));
  wc_Blake2bFinal(&b2b, digest, 64);
  return digest; 
}
\end{lstlisting}
\caption{Function for hashing a byte array using Blake2b}
\label{fig:md-wolf}
\end{figure}

Based on our analysis of the different APIs, we understand that
it is difficult to reuse usage rules across different APIs and
languages. Nonetheless, we found some opportunities to
reuse \csl rules across different JCA providers and within the
Google Tink and Bouncy Castle libraries. These opportunities mostly arise
due to existing security standard recommendations (we can
customize the specifications that address either FIPS or ECrypt recommendations,
for instance), due to the evolution of the API implementations, and due to
the similarity we found among different primitives and primitives'
implementations. We present some examples of these situations in the
remainder of this section.

\begin{mh}
  There is no clear opportunity for reusing the specifications of the
  correct usage of cryptographic libraries across different APIs and languages. 
\end{mh}
%\sk{Removed a line from the box because it was unrelated to this RQ.}

\paragraph*{RQ2: Variability Due to Cryptographic Standards}

Existing technical reports and standards present a
series of recommendations about which cryptographic
algorithms (and respective key configurations) should be used
in applications. These technical reports 
characterize a valuable source of information to
indicate whether a given system is
``secure according to a given standard''.
Moreover, (some) existing cryptographic
APIs (e.g., wolfCrypt and Bouncy Castle) comply to the 
FIPS certifications---and using a certified library
according to the standard recommendations might represent a
competitive advantage for products in specific
domains. For instance, FIPS 140-2 validation is
mandatory for use in the US Federal systems that
collect or store sensitive information.\footnote{https://csrc.nist.rip/groups/STM/cmvp/} 

We found that existing standards introduce a source
of variability in usage specifications. This source
of variability occurs because sets of algorithms (and
algorithm modes) are recommended by some standards, but
not in others. 
%In this section, we highlight this source of variability considering message digests and three different standards that recommend cryptographic usage.
Message digests represent one point in variability as Table~\ref{tabmdfips} shows. All standards mentioned in Section~\ref{subsec:cryptostds} specify secure hash algorithms that may
be used to process a message and produce a condensed
representation (a message digest). However, they do not all recommend the same.

%% \sk{Looking at the table, I wonder if the examples really helps making our case. The only difference appears to be that BSI and ECrypt recommend against SHA-224. Is there no example where the three standards differ more significantly?}

\begin{table}[htb]
%\begin{wraptable}{l}{0.45\linewidth}
\centering
%\begin{small}
  \begin{tabular}{lccc} \toprule
  Algorithm   & FIPS & BSI & ECrypt \\ \midrule
  MD5         & \xmark & \xmark & \xmark \\
  SHA-1       & \xmark & \xmark & \xmark \\
  SHA-224     & \cmark & \xmark & \xmark \\
  SHA-256, 384, 512     & \cmark & \cmark & \cmark \\
  SHA-512/224 & \cmark & \xmark & \xmark \\
  SHA-512/256 & \cmark & \cmark & \cmark \\
  SHA-3/(256, 384, 512)   & \cmark & \cmark & \cmark \\
  Shake128, Shake256    & \cmark & \cmark & \cmark \\
  Whirlpool   & \xmark & \xmark & \cmark \\
  Blake       & \xmark & \xmark & \cmark \\ \bottomrule
  \end{tabular}
%\end{small}
\caption{Recommendations for using different message digest algorithms}
\label{tabmdfips}
\end{table}
%\end{wraptable}

%% None of the standards recommend MD5 and SHA-1 because they are correctly considered insecure, but
%% most of the applications using message digests still use
%% these algorithms (see Section~\ref{sec:msr} \eb{broken ref}) nonetheless.

If we were to encode these standards in \csl, we would need to model them in three distinct rules that nonetheless largely overlap. Let us discuss these rules in more detail. First, consider
the default \csl specification for the \texttt{MessageDigest}
class of the JCA, when considering the default provider 
(Figure~\ref{fig:csl-md-jca}). In this case, the set of
supported algorithms on Line~\ref{line:jca-alg} is limited
to the default algorithms of JCA.

In case we specify aforementioned
standards, we would have to consider using the
Bouncy Castle JCA provider---since the default provider
does not support some of the algorithms in Table~\ref{tabmdfips}
(such as Whirlpool and Blake), and change that
line to consider the recommended algorithms
of each standard, as we show in
Figures \ref{fig:csl-md-jca-variants}. 
In this particular case, it is possible to reuse almost all the
\csl specification of Figure~\ref{fig:csl-md-jca}, changing
only the \texttt{algorithm} constraint based on the
supported standard / technical report. We name this kind
of variability \vsc. % \eb{Better don't use red color.}

\begin{figure}[htb]
\begin{lstlisting}[language=CrySL,numbers=none]
SPEC java.security.MessageDigest
// same definitions of the default JCA MessageDigest specification
CONSTRAINTS
  algorithm in {"SHA-224", "SHA-256", "SHA-384", "SHA-512", "SHA-3", "Shake-128", "Shake-256"};
ENSURES
    digested[out];
\end{lstlisting}

\centering{(a)}

\begin{lstlisting}[language=CrySL,numbers=none]
SPEC java.security.MessageDigest
// same definitions of the default JCA MessageDigest specification
CONSTRAINTS
  algorithm in {"SHA-256", "SHA-384", "SHA-512", "SHA-3", "Shake-128", "Shake-256"};
ENSURES
    digested[out];
\end{lstlisting}

\centering{(b)}

\begin{lstlisting}[language=CrySL,numbers=none]
SPEC java.security.MessageDigest
// same definitions of the default JCA MessageDigest specification
CONSTRAINTS
  algorithm in {"SHA-256", "SHA-384", "SHA-512", "SHA-3", "Shake-128", "Shake-256", "Whirlpool",
  		"Blake2s", "Blake2b"};
ENSURES
    digested[out];
\end{lstlisting}

\centering{(c)}

\caption{\csl rules for the \texttt{MessageDigest} JCA (considering the Bouncy Castle provider and the (a) FIPS recommended algorithms, (b) BSI recommended algorithms, and (c) ECrypt recommended algorithms)}
\label{fig:csl-md-jca-variants}
\end{figure}

Bouncy Castle provides a lightweight API on top of the providers for JCA~\footnote{\url{https://www.bouncycastle.org/}}. Considering the Lightweight Bouncy Castle API,
one is required to write a \csl rule for each
primitive implementation, as shown in Figure~\ref{fig:csl-md-bc} for SHA256 and
SHA512. Instead of one specification for each cryptographic standard (varying
the supported algorithms), there are several
\csl rules for each cryptographic standard (one per supported
primitive implementation). The variability here relates
to the classes that implement the message digest primitives and the
Lightweight Bouncy Castle API implements the individual algorithms
in a distinct class. However, the corresponding
\csl specifications vary only according to the
base class (in the example, \texttt{SHA256Digest} and \texttt{SHA512Digest}).
We name this kind of variability \vbsc.

\begin{mh}
  The specification of the correct usage of cryptographic
  APIs should consider the recommendations of individual
  cryptographic standards. The impact on the specifications
  due to a cryptographic standard depends on the API. 
\end{mh}

% \eb{Also here have some result box(es)?}

\begin{figure}
\begin{minipage}{.45\textwidth}
\begin{lstlisting}[language=CrySL,escapechar=!]
SPEC !\colorbox{white}{SHA256Digest}!

OBJECTS
  byte input;
  byte[] out;
  int outOff;
  
EVENTS
  c : !\colorbox{white}{SHA256Digest();}!
  u : update(input);
  f : doFinal(out, outOff);
	
ORDER
  c, u+, f
		
ENSURES
  digested[out];
\end{lstlisting}
\centering{(a)}
\end{minipage}\hfill
\begin{minipage}{.45\textwidth}
\begin{lstlisting}[language=CrySL,escapechar=!]
SPEC !\colorbox{white}{SHA512Digest}!

OBJECTS
  byte input;
  byte[] out;
  int outOff;
  
EVENTS
  c : !\colorbox{white}{SHA512Digest();}!
  u : update(input);
  f : doFinal(out, outOff);
	
ORDER
  c, u+, f
		
ENSURES
  digested[out];
\end{lstlisting}  
\centering{(b)}
\end{minipage}
\caption{Specification of \csl rules for the message digest classes in
  the Bouncy Castle lightweight API. We will have to elaborate one
  specification for each supported algorithm of a standard / technical
  report.}
\label{fig:csl-md-bc}
%\eb{If you color boxes, maybe use a color one can actually see? :-)}
\end{figure}

\textbf{RQ3: Variability Due to the Evolution of the APIs}

We conduct this study using the approach
introduced in Section~\ref{sec:settings}, to identify cryptographic algorithms introduced/removed
and in turn the breaking changes between two public releases of an API. In this case
we considered three APIs: JCA,
Lightweight Bouncy Castle,
and Google Tink. These APIs already have \csl specifications for them.

Specifically, we mine the
evolution history of 15 releases of the Lightweight
Bouncy Castle (v.1.46 to v.1.60), all available
in the Maven Central Repository.\footnote{https://search.maven.org/} Later we summarize some findings related to the evolution
of the Google Tink and JCA.

The classes that implement
the cryptographic primitives in Lightweight Bouncy Castle implement one of
the existing interfaces declared in the Java package
\texttt{org.bouncycastle.crypto}, including the \texttt{Digest},
\texttt{Mac}, and \texttt{BlockCipher} interfaces.
In the last Bouncy Castle release considered in our
analysis (release 1.60), we identified more than 140 primitive
implementations, among them 45 implementations of
the \texttt{BlockCipher} interface.\footnote{We analyzed these \texttt{BlockCipher} implementations and we found classes that implement cipher
  algorithms (e.g., AES and Blowfish) and cipher modes
  (e.g., CBC and GCM).} Block cipher (45), message
digest (29), message authentication code (18),
and stream cipher (21) are the primitives with
the most algorithm implementations.

\begin{figure}
  \centering{
    \includegraphics[scale=0.8]{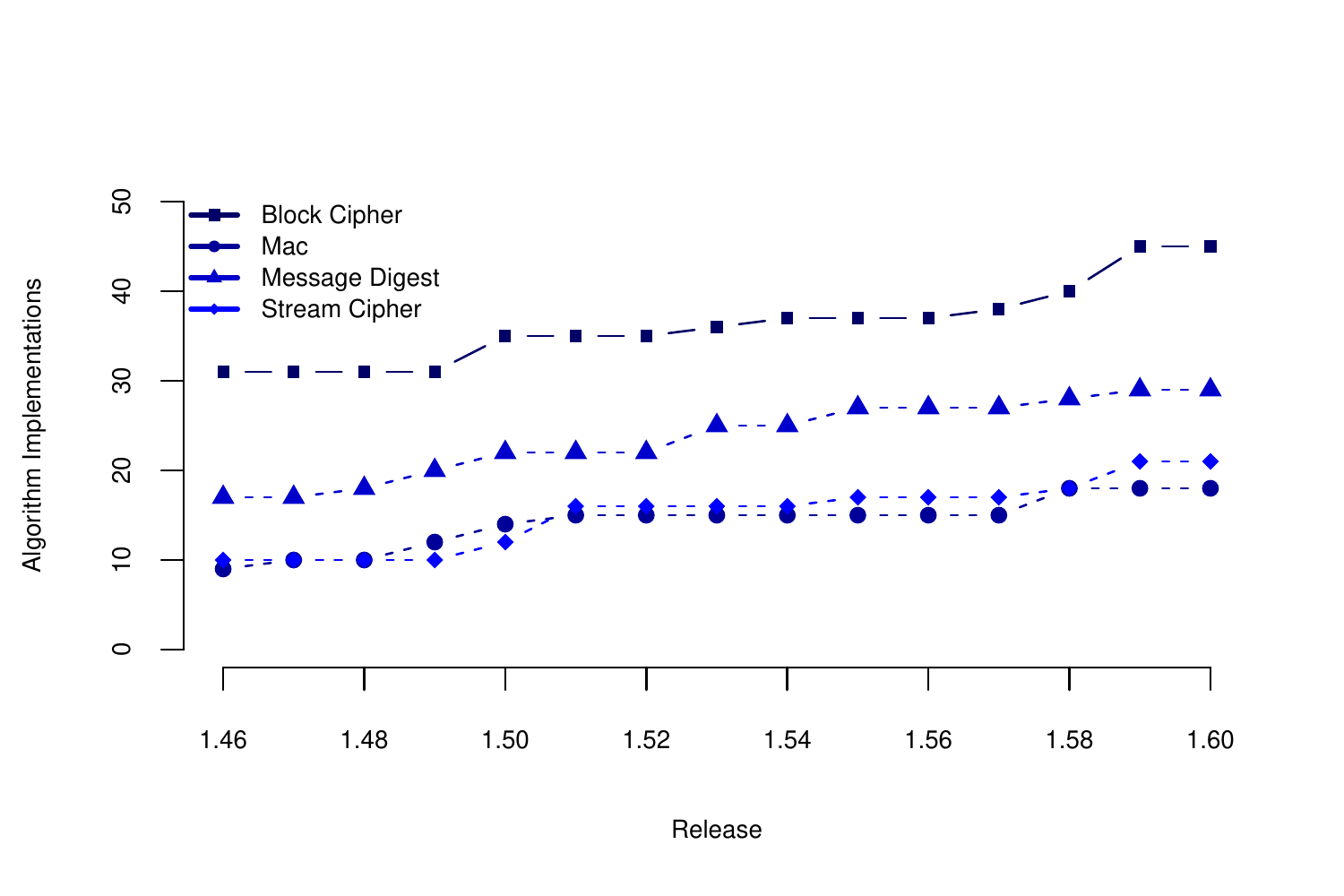}
    }
   \caption{Evolution in the number of algorithm implementations in Bouncy Castle}
   \label{fig:bc-evolution}  
\end{figure}

Figure~\ref{fig:bc-evolution} shows the evolution in the number
of implementations for these primitives. We can see that
almost all releases introduce at least one new primitive
implementation. For instance,
release 1.47 introduced a new implementation of the Mac primitive,
while release 1.59 introduced five new block ciphers,
one new message digest, and three new 
stream ciphers. Only releases 1.52, 1.56, and 1.60 did not introduce
any new such primitive.

The existence of different implementations of a given
primitive has an influence on the specifications of the
correct usage of an API. Consider again the
test case method on Figure~\ref{fig:tc-md}. This example
uses the Bouncy Castle JCA provider (named \texttt{``BC''})
for generating a digest of an input data using the \emph{Blake2b}
algorithm. However, this algorithm was first introduced
in the release \num{1.53} of Bouncy Castle. If one executes this test case
using an earlier release (e.g., 1.51 or 1.52), the test case fails with a
\texttt{NoSuchAlgorithmException}. Therefore,
the \csl specification of Figure~\ref{fig:csl-md-jca-variants}(c)
is not compliant with the releases of Bouncy Castle prior to
\num{1.53}.

% To solve this variability, we can use the
% kind of variability \vsc introduced before. \sk{What do you mean by 'use' here?}

\begin{mh}
  \centering{
    What is considered correct usage of an API depends on the specific
    versions of the API.}
\end{mh}

We also analyzed the changes in the Bouncy Castle
API that might cause an undesired effect on the
client systems~\cite{xavier:saner2017} and identified
\emph{breaking changes}.  Breaking changes include \emph{removing a public method}, \emph{renaming a public method}, \emph{}and
\emph{reducing visibility of a method}. A catalog of these changes
could be found elsewhere~\cite{dig:jsme}.
We only
consider the twelve releases from \num{1.49} until \num{1.60}
because these releases are available in the public Git source
code repository of Bouncy Castle. 

\begin{figure}[tb]
\centering
  \includegraphics[scale=0.6]{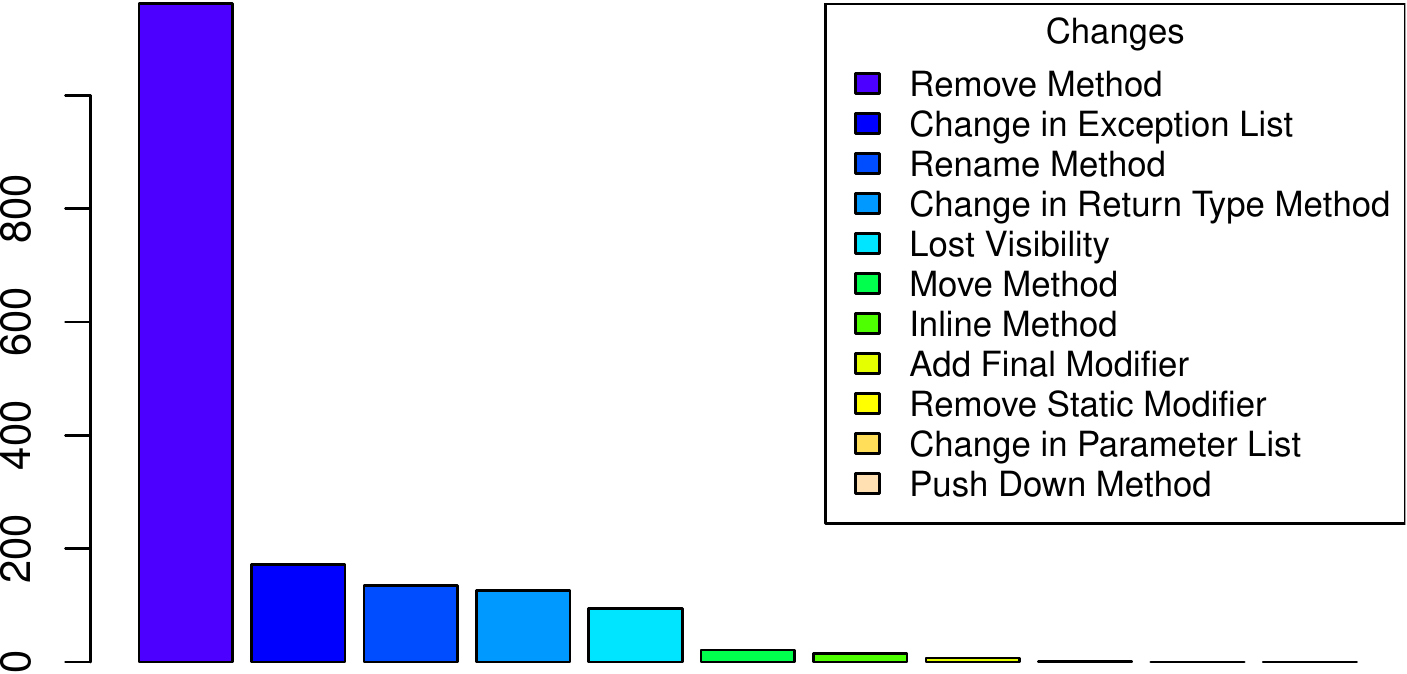}
  \caption{Total number of breaking changes in the Bouncy Castle API}
  \label{fig:bp-brc}
\end{figure}

Using the same approach of previous
works~\cite{api-diff,brito:saner2018,xavier:saner2017}, we found 
\num{1,733} scenarios of breaking changes---considering all pairs
of successive releases. We document them all in Figure~\ref{fig:bp-brc}. In total, we identified \num{1,162} removals of public methods (67\% of all breaking changes). All releases feature at least one. Similarly, all releases change the return type of at least one method. There are a total of
\num{128} occurrences. Other common breaking changes are \emph{change in
exception list} (\num{172} cases), \emph{renaming a public method}
(\num{130} cases), and \emph{reducing visibility of a method} (\num{92} cases).
The remaining 49 breaking fall into other categories.
Four pairs of successive releases contribute with
74.49\% of the breaking changes: 1.58--1.57 (210 cases),
1.57--1.56 (364 cases), 1.51--1.50 (389 cases),
and 1.50--1.49 (328 cases). We did not find any evidence
that one specific release accounts for a major redesign of
the Bouncy Castle library.

Based on these numbers, one might conclude
that, despite its long history, Bouncy Castle 
is an unstable library. This conclusion would be true if
developers depended on the public interfaces of the
classes that present breaking changes. However,
when we consider only the Java interfaces that define
the contract of cryptographic primitives (such as the
\texttt{Digest} and \texttt{BlockCipher} interfaces),
we found that the Bouncy Castle library is quite stable.
Considering all releases, we only identified 34 breaking
changes (12 occurrences of \emph{removing a public method}, 9
occurrences of \emph{changing the return type of
a public method},
7 occurrences of \emph{renaming a public method}, 4
occurrences of \emph{reducing visibility of a method}, and 2
occurrences of \emph{changing the exception list of a method}). Yet, we do not know whether or
not developers only rely on these ``high level''
interfaces. 
%\sk{Given their significance for this section and even more so for the design of \mcsl, shouldn't we discuss these 34 cases in more detail here?}\kn{we are already short of space :( }

\begin{mh}
  We found \num{1,733} breaking changes along 11 public releases
  of Bouncy Castles. However, considering the core interfaces
  of the library, we only found 34 breaking changes that
  might also induce changes on \csl specifications. 
\end{mh}

Method updates like renaming/removing/adding methods requires changes to the event section of \csl specifications. We name this variability as \vces.

%% \begin{table}[thb]
%%   \caption{Evolution in the number of primitive implementations in Bouncy Castle}
%%   \label{tab:bc-evolution}
%% \begin{tabular}{cccccc} \toprule
%% Release & Block Cipher  & Mac  & Message Digest & Stream Cipher \\ \midrule       
%% 1.46    & 31            & 9    & 17             & 10   \\ 
%% 1.47    & 31            & 10   & 17             & 10   \\
%% 1.48    & 31            & 10   & 18             & 10   \\
%% 1.49    & 31            & 12   & 20             & 10   \\
%% 1.50    & 35            & 14   & 22             & 12   \\
%% 1.51    & 35            & 15   & 22             & 16   \\
%% 1.52    & 35            & 15   & 22             & 16   \\
%% 1.53    & 36            & 15   & 25             & 16   \\
%% 1.54    & 37            & 15   & 25             & 16   \\
%% 1.55    & 37            & 15   & 27             & 17   \\
%% 1.56    & 37            & 15   & 27             & 17   \\
%% 1.57    & 38            & 15   & 27             & 17   \\
%% 1.58    & 40            & 18   & 28             & 18   \\
%% 1.59    & 45            & 18   & 29             & 21   \\
%% 1.60    & 45            & 18   & 29             & 21   \\ \bottomrule
%% \end{tabular}
%% \eb{Turn this into a line chart}
%% \end{table}

%% We considered five releases of Google Tink that are publicly available
%% at the Maven Central Repository. Google Tink presents a design quite different
%% from Bouncy Castle, and the different algorithms of a given primitive
%% are not implemented using a type hierarchy. \sk{It is way too late to introduce this sort of stuff here. This section is about variability, not about design decisions of Google Tink or other APIs.} That is, the API has only one
%% implementation of the Authenticated Encryption with Associated Data (AEAD)
%% primitive, for instance, which could be customized through specific key templates.
%% Based on this design, to instantiate an AEAD object,
%% it is necessary to build a new key template or reuse one of the existing
%% key templates for a given primitive---we believe that this second approach
%% is less error-prone and more suitable for most use cases. To make this
%% discussion more clear, Figure~\ref{fig:gt-example} shows
%% an example of the use of the AEAD primitive configured with an
%% existing key template (\texttt{AES128\_GCM}), based on the
%% AES algorithm and the GCM mode of operation.

%% \begin{figure}[htb]
%% \begin{lstlisting}[language=Java]
%% KeysetHandle k = KeysetHandle.generateNew(AeadKeyTemplates.AES128_GCM);
%% Aead aead = keysetHandle.getPrimitive(k);
%% byte[] ciphertext = aead.encrypt(plaintext, aad);
%% \end{lstlisting}
%% \caption{Example of use of the Google Tink AEAD primitive}
%% \label{fig:gt-example}
%% \end{figure}

We further investigated whether the
Google Tink library is more stable: the public interfaces of
the classes are almost unchanged between the release 1.0.0
(published in September 2017) and the release 1.2.1 (published
in November 2018). During this period,
we found 50 breaking changes---43 from version 1.0.0 to
version 1.1.0, which might indicate a slight revision on the first
design of the library. Between these first initial releases, we
identified 14 removed methods. Nevertheless, the most critical change
regarding \csl specifications was the introduction
of the Deterministic AEAD algorithm on version 1.1.0. %Prior to this release, one must not use the set of \csl specifications related to this algorithm. 
This type of variability is modular and involves only the selection of a set of \csl specifications (hereafter referred to as \vmsr). Although not common,
we also identified some variability due to the key templates available
across the different versions of Google Tink. The
current \csl specifications for Google Tink can deal with the
introduction of key templates that modify the events in the specifications (using the \vces strategy). % \sk{What about the other 48 changes?}

Finally, we also considered the evolution of 
JCA from Java 4 to Java 9. To this end, we analyzed the classes
related to the cryptographic primitives available in
three standard Java libraries: \texttt{rt.jar}, \texttt{jce.jar}, and
\texttt{sunjce\_provider.jar},
considering official releases of the Java language.
This API is highly stable as it is based on an
official Java specification. For instance, 
the public class interfaces of the JCA
do not present any breaking change, and from
Java 5 (2005) to Java 9 (2017) the interface of the
\texttt{java.security.MessageDigest} class did not change.
In Java 7, three additional methods that can be
used for ciphering a text with \emph{additional authentication
data} (AAD) were introduced in the class
\texttt{javax.crypto.Cipher}.
Although the APIs are stable, new primitive algorithms
have been introduced along these versions. For instance,
eight new ciphers and six new MAC algorithms have been
introduced in the JCA, from Java 4 to Java 9.
% In conclusion, the JCA does display similar sorts of variability as BouncyCastle.

%% \subsection{Reuse Opportunities on APIs}

%% {\color{red} Pending \ldots Here we will detail some
%%   reuse opportunities we identified in both Lightweight
%%  Bouncy Castle and Google Tink.}

\section{\mcsl}\label{sec:design}

\subsection{Design and Implementation Procedures}

We used the outcomes of our domain analysis to
design and implement \mcsl. \mcsl provides means for the
systematic reuse of \csl specifications. To this end,
\mcsl allows the specification of \csl rules enriched
with variation points (such as meta-variables and
type parameters) and {\bf refinement} operations that
solve these variation points for a given {\bf configuration}
(e.g., version of an API or
platform, security standard, and so on). \mcsl generates a
set of \csl rules tailored for a given configuration.

We implemented \mcsl using Rascal-MPL~\cite{klint:scam2009}.
%and the
%recommendations of existing patterns for
%\emph{language implementation}~\cite{lip-book} (e.g., External Tree Visitor and
%Generating DSLs with Templates) and the
%pipeline architectural style~\cite{eip-book}.
One of the main design
decisions was to implement
three distinct languages: one for abstract \csl specifications (i.e., \csl
with variation points), one for \csl refinements,
and one for representing a configuration model. The configuration
model states a set 
of abstract \csl specifications and refinements. We use a program-generator approach to combine instances
of the refinement and configuration languages, and to output regular \csl specifications. These regular specifications can directly be used with CogniCrypt's infrastructure for \csl specifications. 
The following set of high-level requirements guided
the design of \mcsl. 

\begin{enumerate}
\item \mcsl follows a \emph{meta-programming} approach: we write \mcsl specifications and 
  generate regular \csl specifications from them.
	Using this design allows us to preserve all
	\ca infrastructure. 

      \item \mcsl should
	support the sources of
	variability discussed in the previous
	section, so that we can generate
	\csl rules for different standards and
	versions of the APIs. 
        
      \item \mcsl should also favor
	reuse among specifications of the same
	API, reducing the effort in the case that an
	API supports many algorithms (as for instance
	Bouncy Castle). 
\end{enumerate}

%Commenting the following as it is 1) redundant and 2) Rascal is known to the community already and we dont need to justify why Rascal is good
%In view of the above requirements, we implemented
%a first version of \mcsl using the \rascal
%meta programming language~\cite{klint:scam2009}.
%\rascal\sk{I would generally opt to explain methodology only once per section.} presents first class constructs for
%program analysis and manipulation, including
%pattern-matching on concrete syntax and the
%implementation of the visitor pattern as a
%programming language construct. It is also possible to
%reuse grammar definitions, in a similar fashion
%that developers reuse program modules. We considered
%this feature valuable in our design, since we have \sk{Why do we *have to*?}
%to reuse grammar definitions of an extended version
%of the \csl language in other grammar modules
%of \mcsl. Our previous experience 
%also motivates the decision for 
%implementing \mcsl using Rascal-MPL.

\subsection{High-level Architecture}

\begin{figure}[htb]
\centering
 \includegraphics[scale=0.45]{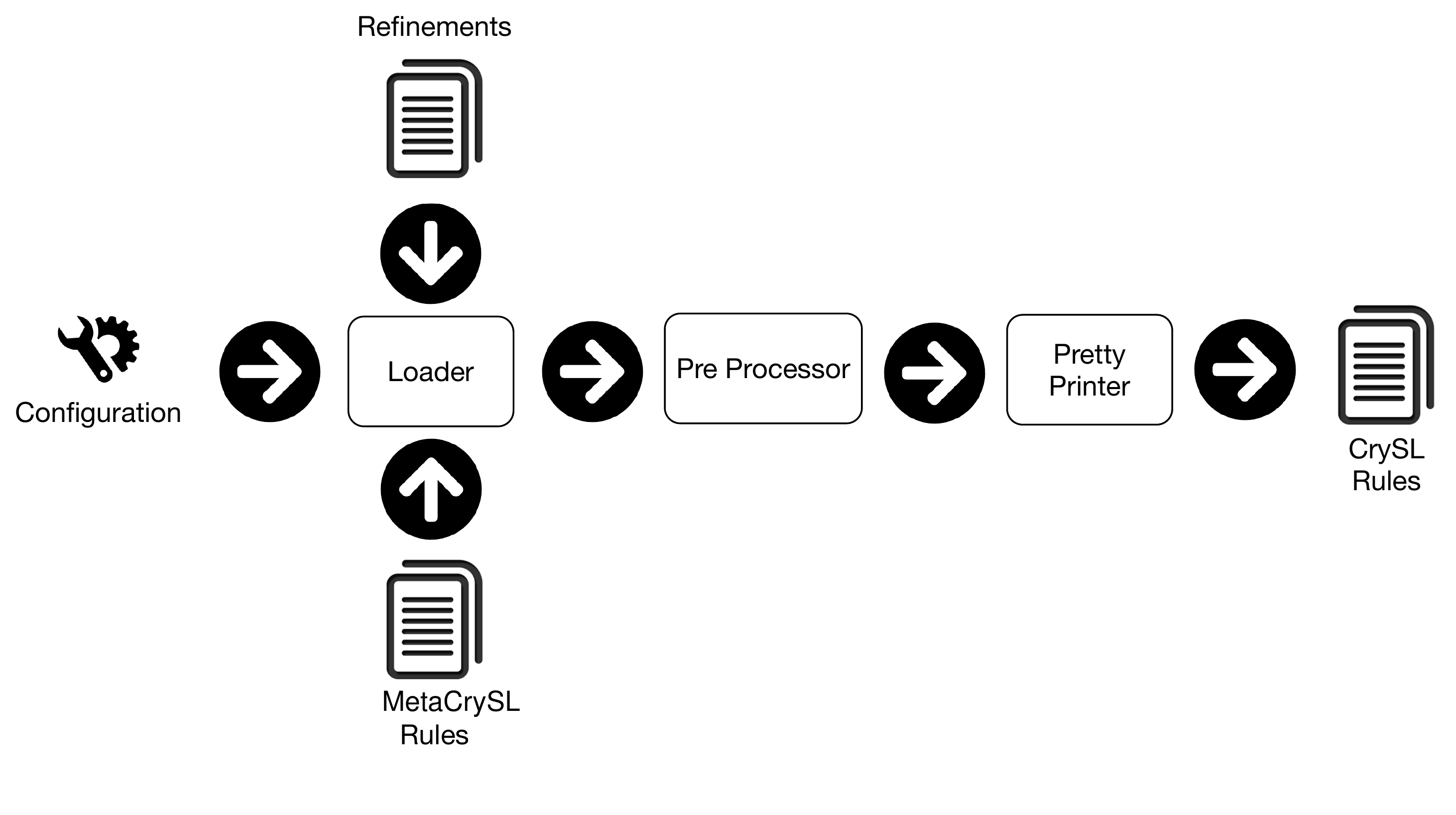}
  \caption{High-level architecture of \mcsl}
  \label{fig:architecture}
\end{figure}

Figure~\ref{fig:architecture} shows the 
architecture of \mcsl, which
follows a multi-staged pipeline for language
processing~\cite{lip-book}, where
a module loads a \mcsl \texttt{configuration} that
specifies a set of extended \csl specifications and
a set of \texttt{refinements} that should be used during the
building process of a specific set of \csl rules. After
that, the \emph{Loader} module parses the sets of 
extended \csl and refinement files, generating
abstract representations of these languages as
instances of \rascal algebraic data types (in the following
sections we detail these languages). 
The \emph{Preprocessor} module manipulates these
instances executing the refinement operations, 
using visitors for program transformations. That is, the
\emph{Preprocessor} solves \mcsl variability and
generates an abstract representation of \csl rules. Finally, the
\emph{Pretty Printer} module outputs regular
\csl specification files. 

\subsection{Abstract \csl Language}

Abstract \csl is an extension of
the \csl language that allows cryptographic
specialists to write variation points on
the \csl rules, for instance, in terms of \emph{meta-variables} and
type parameters.  Figure~\ref{fig:ecsl-md} shows an
example of an instance of the abstract
\csl language, modelling variability on
\csl rules for the JCA \texttt{MessageDigest} class.
The main source of variability in
this case relates to the sets of algorithms
that might change due to a specific standard or
version of the provider implementation (recall the
specifications in Figure~\ref{fig:csl-md-jca-variants}).
The abstract \csl rule of 
Figure~\ref{fig:ecsl-md}
introduces the concept of \emph{meta-variables},
which are bound during the derivation process
of \csl rules. In the example, we can bind
the meta-variable \texttt{\$AlgSet} to
the sets of algorithms supported by a given
standard (e.g., FIPS, EuroCrypt, or BSI) or
specific version of an API. 

\begin{figure}[htb]
\begin{lstlisting}[language=CrySL,numbers=none]
SPEC java.security.MessageDigest
// same definitions of the default JCA MessageDigest specification
CONSTRAINTS
  algorithm in $AlgSet;
ENSURES
    digested[out];
\end{lstlisting}
\caption{Use of meta-variables to deal with \vsc}
\label{fig:ecsl-md}
\end{figure}

To deal with \vbsc,
we use template-based type parameters, similarly to
the mechanism of type expansion supported by C++ templates. As such, 
when solving this type of variability,
we actually generate different copies of
a \csl rule, one for each
concrete type that appears in the refinement specifications.

Using the abstract specification of
Figure~\ref{fig:ecsl-tp}, we generated six \csl rules
for different Google Tink primitive's
implementations. We also used the same strategy to factor out 
existing \csl rules for the message digest primitive
of the Lightweight Bouncy Castle API. Each one of
these \csl rules has about 40 lines of code.
Using \mcsl, we were able to specify each
variant using 4 lines of code (three lines
of refinements and one line of the configuration
language). 

\begin{figure}[htb]
\begin{lstlisting}[language=CrySL, numbers=none]
ABSTRACT SPEC AbstractFactory<T>
OBJECTS
  com.google.crypto.tink.KeysetHandle ksh;
  <T> primitive; 
EVENTS
  gp : primitive = getPrimitive(ksh);
ORDER
  gp
REQUIRES 
  generatedKeySet[ksh]; 
ENSURES
  setPrimitive[primitive]; 
\end{lstlisting}
\caption{Use of type parameters to deal with \vbsc}
\label{fig:ecsl-tp}
\end{figure}

\subsection{Refinement Language}

Our refinement language allows cryptographic experts
to specify transformations on the \mcsl rules, to
solve variation points. Considering the discussion of
the previous section, \mcsl supports
two types of syntactic variation points: meta-variables and
type parameters. In addition, it is also possible
to introduce new events (and events aggregates), to introduce new
constraints, and to replace the events' order of a \mcsl specification.
The refinement language expects a
base specification and a list of refinement
elements. 

The current implementation of \mcsl supports different
refinement transformations, for instance, transformations that
support the kinds of variability discussed in the previous
section:

\begin{itemize}
\item {\bf Define literal set} binds a meta-variable to a literal
  set, such as \texttt{SHA512, Blake2b, Blake2s}. We use this
  transformation to solve \vsc.

\item {\bf Define qualified type} binds a fully qualified type
  to a type parameter of a \mcsl specification. This
  transformation solves \vbsc. 

\item {\bf Add new event} introduces a new event into
  a \mcsl specification. We use this transformation to solve
  \vces. Similarly, the refinement language also supports
  operations to add (remove or update) constraints and
  requires/ensures clauses. 
\end{itemize}

Figure~\ref{fig:refinements-bc-provider}
shows two examples of  \emph{refinement
  specifications}. The first  
\emph{refines} the \texttt{MessageDigest}
\csl specification of Figure~\ref{fig:ecsl-md}, binding the meta-variable
\texttt{AlgSet} to a set of message digest
algorithms supported by the Bouncy Castle JCA
provider. The second refinement specification
\texttt{KeyGenerator} also defines a set of
algorithms supported by the Bouncy
Castle JCA provider
and also introduces a new constraint which refers to
two variables of the base specification (not
illustrated in this paper): \texttt{alg}
and \texttt{keySize}. The constraint states
that if $alg = AES$, the
variable \texttt{keySize} must be a value in
the set $\{128, 192, 256\}$. 

\begin{figure}
\begin{lstlisting}[language=CrySL]
SPEC MessageDigest REFINES java.security.MessageDigest {
  define AlgSet = {"Blake2s", "Blake2b", "GOST-3411", "SHA-256", "SHA-384",
                   "SHA-512", "Whirlpool"};
}
SPEC KeyGenerator REFINES javax.crypto.KeyGenerator {
  define AlgSet = {"AES", "BLOWFISH", "HmacSHA256", "HmacSHA384", "HmacSHA512",
                   "RIJNDAEL", "Serpent"};
  add constraint alg in {"AES"} => keySize in {128, 192, 256};
}
\end{lstlisting}
\caption{Example of refinement specifications for the Bouncy Castle JCA Provider.}
\label{fig:refinements-bc-provider}
\end{figure}

Figure~\ref{fig:refinements-tp} shows a set of
refinement specifications that are used for
generating \csl rules for different message
digest algorithms supported by the 
Lightweight Bouncy Castle API. Each refinement
specification generates a different \csl specification,
binding a type parameter with the full qualified name
of a class that implements a message digest algorithm.
In this scenario, we are able to solve all variability
using only type parameters, and thus the body of the
refinement specifications is empty.

\begin{figure}
\begin{lstlisting}[language=CrySL]
SPEC SHA256 REFINES
     Digest<org.bouncycastle.crypto.digests.SHA256Digest>;
SPEC SHA384 REFINES
     Digest<org.bouncycastle.crypto.digests.SHA384Digest>;
SPEC SHA512 REFINES
     Digest<org.bouncycastle.crypto.digests.SHA512Digest>;
SPEC SHA512t REFINES
     Digest<org.bouncycastle.crypto.digests.SHA512tDigest>;
\end{lstlisting}
\caption{Example of refinements that bind a type parameter for the
  set of message digest specifications for the Lightweight Bouncy
  Castle API.}
\label{fig:refinements-tp}
\end{figure}

It might be necessary to add further refinement transformations
in the future. To implement a new transformation,
one would have to modify three \rascal modules,
being necessary to specify the concrete and abstract syntax
of the transformation in the refinement language
and to implement a new function with the expected behavior of the
transformation (\texttt{Preprocessor} module).
In case one needs
to introduce a new syntactic \csl variation point, this is possible by modifying the abstract and concrete syntax of the abstract \csl language.
We have already implemented six transformations, each one having around ten lines of code.

\subsection{\mcsl Configurations}

We use a configuration language to specify the \mcsl
\emph{building process}. Figure~\ref{fig:config-sample} shows an example, which states
the base path where the \mcsl implementation
could find the specifications and refinements (Line~2), 
the output path of
the resulting \csl specifications (Line~3), and the sets of
abstract \csl rules and refinements that should
be considered during the building process (Lines~4--7).
In the example, all \csl rules reside in the \texttt{base} directory. One may also
specify individual rules instead of a directory. 
The specification of a building process
allows cryptographic experts to reuse the same
specifications and refinements in different configurations.
That is, from the same set of Lightweight Bouncy
Castle specifications and refinements, we can create different configurations and generate
distinct sets of \csl rules. For this flexibility, we opted for such a
\emph{configuration language} instead of
a \emph{convention-based} mechanism.

\begin{figure}
\begin{lstlisting}[language=configuration, numbers = left]
config android25plus {
  src = MetaCrySL/samples/jca/base/; 
  out = MetaCrySL/samples/jca/android/target/research/25plus/;
  load spec base/;	
  load refinement android-bsi/01plus/;
  load refinement android-bsi/10plus/;
  load refinement android-bsi/1025/;
}
\end{lstlisting}  
\caption{Example of a configuration that specifies the rules and refinements target to the
version 25 of Android}
\label{fig:config-sample}
\end{figure}

\section{Empirical Assessment of \mcsl}\label{sec:eval}

  The {\bf goal} of this empirical assessment is to
  understand the implications of \mcsl in modularizing the
  specifications of the correct usage of the JCA API for
  Android, and thereby evaluating \mcsl along the lines of \emph{compactness}.
  Additionally, we also use the empirical assessment to investigate
  whether or not \mcsl generates correct \csl specifications, focusing
  on the \emph{correctness} dimension. Accordingly,
  we answer the following research {\bf questions} in this empirical assessment,
  where RQ4 and RQ5 relate to \emph{compactness} and
  RQ6 explore the \emph{correctness} perspective:

  % \sk{Why do we not format them as proper RQs?}
  % \eb{Do so! Define these at the beginning of Section 5.}

  \begin{enumerate}[RQ1] \setcounter{enumi}{3}
    \item \rqcode 
    \item \rqdup 
    \item \rqviolations 
\end{enumerate} 

  Answering \textbf{RQ4} and \textbf{RQ5} allows us to
  quantify the main expected benefit of \mcsl: modularizing
  families of \csl specifications with the aim of
  specification reuse. Answering \textbf{RQ6} allows us (a) to contrast the difference in the
  number of reported API misuses when evaluating
  programs using different \mcsl configurations and
  (b) to check the correctness of our approach for
  generating \csl rules (since \ca will reject any
  invalid \csl rule). 
In this assessment, we used \mcsl to modularize a family of \csl
specifications supporting different {\bf versions} of the
Android platform and three {\bf sets of cryptographic recommendations}:

\begin{itemize}
\item {\bf Android Base recommendations}:
  constrains the algorithms that should be used for
  each version of the Android platform, as detailed
  in the Android Cryptography Guide specification.\footnote{Android Cryptography Guide: \url{https://developer.android.com/guide/topics/security/cryptography}}
  
\item {\bf Android BSI standard recommendations}: constrains
  the algorithms considering the BSI standard and the
  set of Android Base recommendations. The set of Android Base
  recommendations must be considered because not all BSI
  recommended algorithms are available in every version of the Android
  platform. 
  
\item {\bf Android CogniCrypt recommendations}: constrains
  the algorithms according to the current \csl specifications
  from the CogniCrypt project and the set of Android
  Base recommendations. The set of Android Base
  recommendations must be considered because not all CogniCrypt
  recommended algorithms are available in every version of the Android
  platform. 

\end{itemize}

Specifying the correct usage of the JCA for Android
is an interesting scenario for using
\mcsl, in particular because 
the correct usage of cryptography in
Android depends on the version of the Android platform.
Moreover, to answer our research question RQ6, this decision allows us to leverage the
same dataset of Android applications that was previously
used to empirically assess \csl~\cite{kruger:tse2019}. This dataset
comprises \num{8136} Android applications, though
we could not collect the output of the \ca 
for at least
one configuration in a subset comprising \num{507} of these Android apps. For this reason,
we consider a smaller set of \num{7629} Android apps.
From our \mcsl specifications, we can generate hundreds
of configurations. Since it is computationally expensive to
run \ca on a dataset with thousands of Android apps,
we decided to conduct our assessment with the nine configurations
shown in Table~\ref{tab:assessment-sets}. Each
configuration supports all cryptographic primitives (JCA supports
32 primitives in total, including Block Cipher and Message Digest),
one of three distinct ranges of versions of the Android platform (01 -- 08, 01 -- 16,
01 -- 28), and one of the cryptographic recommendations.
  
  \begin{table}
    \begin{scriptsize}
    \begin{tabular}{llcl}\toprule
      Config. Id & Primitives  & Android Platform Version & Crypto Standard \\ \midrule
      C01 & All primitives & 01 -- 08  & Android Base recommendations \\
      C02 & All primitives & 01 -- 16  & Android Base recommendations \\
      C03 & All primitives & 01 -- 28  & Android Base recommendations \\
      C04 & All primitives & 01 -- 08  & Android BSI Standard recommendations \\
      C05 & All primitives & 01 -- 16  & Android BSI Standard recommendations \\
      C06 & All primitives & 01 -- 28  & Android BSI Standard recommendations \\
      C07 & All primitives & 01 -- 08  & Android CogniCrypt recommendations \\
      C08 & All primitives & 01 -- 16  & Android CogniCrypt recommendations \\
      C09 & All primitives & 01 -- 28  & Android CogniCrypt recommendations \\ \bottomrule  
    \end{tabular}
    \end{scriptsize}
    \caption{Sets of cryptographic rules considered in our study}
    \label{tab:assessment-sets}
  \end{table}
  
  We answer research questions \textbf{RQ4}, \textbf{RQ5}, and \textbf{RQ6} through the use of metrics.
  For \textbf{RQ4} we compute (a) the total number of lines in \mcsl 
  necessary to specify the sets of configurations
  of Table~\ref{tab:assessment-sets} and (b) the resulting
  lines of specifications in \csl that we generate using the \mcsl
  specifications. We then compute how many lines of specification text
  we save using \mcsl. For \textbf{RQ5} we estimate the total number of duplication
  in the \mcsl specifications, as well as in the  generated \csl rules. 
  We answer \textbf{RQ6} using
  the \emph{total number of rule violations} that \ca finds in the dataset of Android applications when using each distinct set of \csl rule configurations.

\subsection{RQ4: \rqcode}
  
\begin{figure}[htb]
\centering
    \includegraphics[scale=0.8]{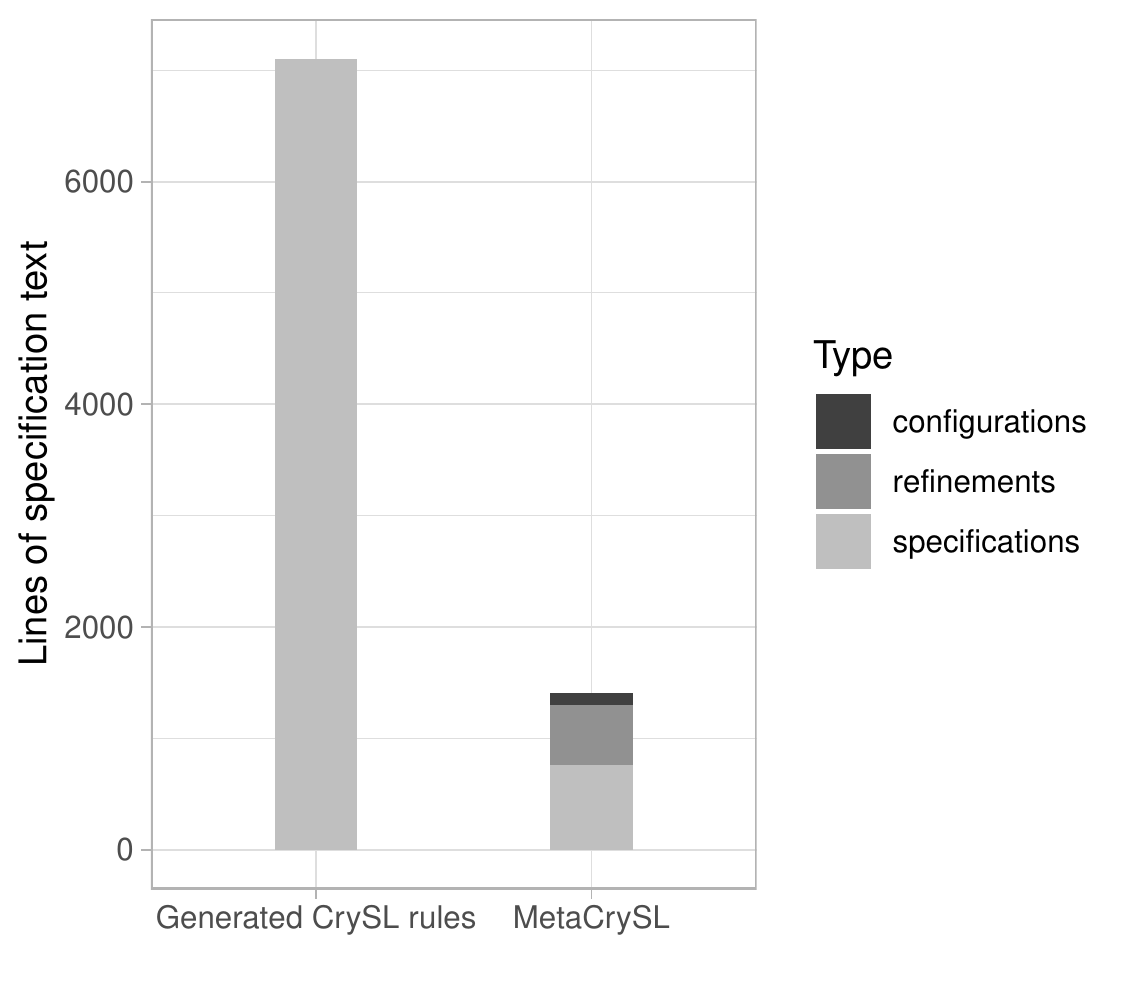}
    \caption{The total number of lines of code necessary to
      specify the nine configurations in \mcsl (including \mcsl
      specifications, \mcsl refinements, and \mcsl configurations) and the
     total number of lines of \csl code generated.}
    \label{fig:bp-sloc}
  \end{figure}

  In \textbf{RQ4}, we  investigate
  the benefits of using \mcsl w.r.t.\ removing the redundant 
  code that one would write when specifying the sets of \csl rules
  describing the correct usage of cryptographic APIs---tailored
  to the nine configurations in Table~\ref{tab:assessment-sets}.
  Figure~\ref{fig:bp-sloc}
  summarizes the total number of lines needed to
  write the \mcsl specifications, refinements, and configurations as well as
  the total number of lines of specifications generated by \mcsl and that could
  be used to execute \ca with the distinct configurations. 
  In this case study,
  we wrote \num{1407} lines in \mcsl (762 lines of \mcsl specifications,
  540 lines of \mcsl refinements, and 105 lines of \mcsl configurations), and
  generated \num{7105} lines of \csl rules for
  those configurations, saving 80\%  of lines.

  %% RB: Threats to validity is already discussing this issue.
  
  % \eb{Is this truly a valid comparison? I wonder if the generated rules could
  % contain boilerplate code that would not have arisen had these rules been
  % written manually.}

  \begin{mh}
    \mcsl removed 80\%  of the redundancy induced when writing
    all the \csl rules tailored to the specific configurations
    considered in our study.  
  \end{mh}

\begin{figure}[htb]
    \includegraphics[trim=0 20 0 30]{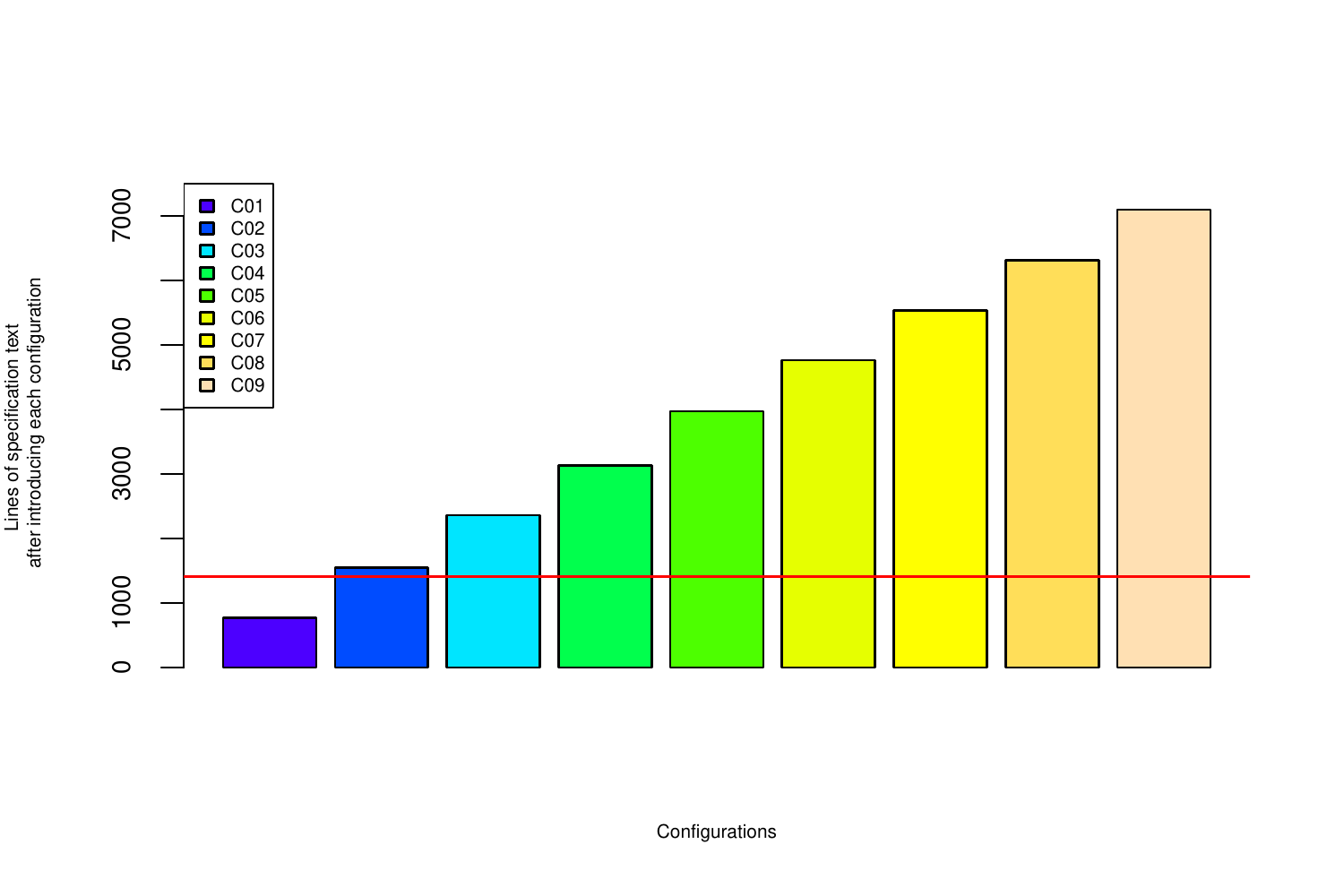}
    \caption{Evolution of the total lines of generated \csl specification text
      after introducing each configuration. The red line corresponds to the
      total number of lines of \mcsl used to generate the configurations}
    \label{fig:accumulator} 
\end{figure}

  The \mcsl payoff tends to increase when defining
  new configurations, since one would then generate further instances 
  of \csl from the same set of \mcsl rules and refinements.
  Figure~\ref{fig:accumulator} shows how many lines of \csl
  specification we generate after introducing each configuration in
  Table~\ref{tab:assessment-sets}.
  In terms of lines of specification text, we achieve a
  payoff after generating the second
  configuration (C02). 

\subsection{RQ5: \rqdup}

In total, there were $188$ files (including refinements and configurations) of base \mcsl specifications for the JCA use in Android. These files contained $1407$ lines of specifications, out of which $633$ lines were duplicates, resulting out of $156$ individual lines. In comparison, the corresponding \csl specifications for three families of Android configurations (BSI, CogniCrypt, Base) each comprising specifications for three versions (0108, 0116, 25plus) contributed to \num{7105} lines of specifications spread across $288$ files. Out of these, \num{5579} lines of specifications were duplicates resulting out of 546 unique lines.

  \begin{mh}
  The amount of duplicate lines of specifications for a family of \csl specifications is \num{5579} in comparison to \num{633} for \mcsl specifications for the same family (\num{11.34}\%).  
  \end{mh}

  %% \eb{Can we say something about where duplication still exists? I'd find that an interesting aspect.}\rb{This is an interesting point. In fact, we could have removed most of the
  %%   duplications in the \mcsl specifications, by factoring out the similarities within the
  %% refinements for Android-BSI and Android-CC.}

  Most of the duplication in \mcsl arises because we specified all
  \mcsl refinements for the three families of Android configurations
  (BSI, CogniCrypt, and Base) which could be prevented by writing carefully crafted refinements. Specifically, out of the $1407$ lines of specifications, only $97$ lines and $85$ lines of duplicates resulted from the base \mcsl specifications and configurations--- $451$ of the $633$ duplicates resulted from refinements for individual versions.

  \subsection{RQ6: \rqviolations}
  
  Our research question RQ6 explores the
  results of \ca for the nine configurations (C01 -- C09). We concentrate our
  analysis on the violations related to the \texttt{CONSTRAINTS} section
  of \csl rules, mostly because cryptographic standards do not address other sections.
  Table~\ref{tab:mcl-results-rq2}
  summarizes the results of the analysis, showing the number of
  Android apps using the JCA APIs, the number of Android apps using
  the JCA APIs incorrectly (i.e., presenting at least one misuse), the
  rate of vulnerable Android apps (calculated using the previous two), and the total number of
  violations.
  The results of \ca reveal a significant number of
  apps with at least one JCA API misuse in all configurations---more than five percent 
  of the apps present at least one misuse in the
  more permissive Android Base sets of recommendations. This number jumps to more than 45\% when considering the BSI or the CogniCrypt recommendations. 

 % latex table generated in R 3.6.1 by xtable 1.8-4 package
 % Thu Apr  2 14:15:00 2020
 \begin{table}[ht]
   \centering
   \begin{small}
 \begin{tabular}{lrrr|r}
   \toprule
 Configuration  & Apps & Apps Presenting  & (Rate \%) &  \# Violations \\ 
                & Using JCA  & Misuse & & \\
   \midrule
Android Base 0108    & 6,714 & 545 & 8.12 & 1,083 \\ 
Android Base 0116    & 6,714 & 395 & 5.88 & 1,224 \\ 
Android Base 25plus  & 6,714 & 386 & 5.75 & 830 \\ 
Android BSI  0108    & 6,714 & 3,184 & 47.42 & 9,089 \\ 
Android BSI  0116    & 6,714 & 3,155 & 46.99 & 8,905 \\ 
Android BSI  25plus  & 6,714 & 3,155 & 46.99 & 8,873 \\ 
Android CogniCrypt 0108 & 6,714 & 3,261 & 48.57 & 9,077 \\ 
Android CogniCrypt 0116 & 6,714 & 3,260 & 48.56 & 8,975 \\ 
Android CogniCrypt 25plus  & 6,714 & 3,256 & 48.50 & 8,945 \\ 
\bottomrule
 \end{tabular}
 \end{small}
 \caption{Summary of the findings of \ca for the
   different CrySL configurations.}
 %These results only
 %  consider the \texttt{constraints} section of \csl and
 %  discards common libraries}
 \label{tab:mcl-results-rq2}
%  \eb{I think we should switch the last two columns as the rate is based on the other numbers.}
 \end{table}

%% Android Base & 0108         & 6714 & 850 & 12.66 & 1415 \\
%% Android Base & 0116         & 6714 & 396 & 5.90 & 1225 \\
%% Android BASE & 25plus       & 6714 & 387 & 5.76 & 831 \\ 
%% Android BSI  & 0108         & 6714 & 6600 & 98.30 & 21459 \\ 
%% Android BSI  & 0116         & 6714 & 6586 & 98.09 & 20944 \\ 
%% Android BSI  & 25plus       & 6714 & 6586 & 98.09 & 20912 \\ 
%% Android CogniCrypt & 0108   & 6714 & 6642 & 98.93 & 19657 \\ 
%% Android CogniCrypt & 0116   & 6714 & 6641 & 98.91 & 19555 \\ 
%% Android CogniCrypt & 25plus & 6714 & 6639 & 98.88 & 19525 \\
 
The total number of violations when considering the set of Android
 Base recommendations is substantially smaller than the total number of violation found using the other configurations (Android-BSI and Android-CC configurations) and most of the violations in the Android Base configurations relate to the \emph{Cipher}
   primitive. For instance, when one only considers the ``Android Base 25plus'' configuration,
 548 out of 830 violations are either due to the use of an insecure cipher algorithm (such as
 DES or DESede) or due to the use of an insecure algorithm/mode/padding
 configuration (e.g., AES/CBC/NoPadding). This changes when one considers the other sets
 of recommendations (from Android BSI and Android CogniCrypt). There, most of the violations
 relate to the \emph{Message Digest} primitive. For instance, considering the
 Android BSI \texttt{0116} configuration, one finds 6,272 violations due to insecure message-digest algorithms (e.g., MD5 and SHA-1)---this corresponds
 to 70.43\% of all violations one finds with this particular configuration.

 Regarding the differences between the Android BSI and
 Android CogniCrypt families of \csl rules we found
 some modes of operations that are not mentioned in the
 BSI standard (e.g., RSA/ECB/PKCS1Padding) but that are
 considered secure and recommended in CogniCrypt.
 The \emph{Message Authentication Code} (MAC)
 primitive also brings differences in the number of
 violations when comparing the BSI and CogniCrypt recommendations. Actually,
 the BSI standard makes clear that the HMAC scheme should only be
 used with the SHA-2 or SHA-3 families of hash functions, though
 the algorithms \texttt{HmacMD5} and \texttt{HmacSHA1} are
 allowed by the CogniCrypt configurations.  

 % \eb{When they are not mentioned what does
 % that mean for the CrySL spec? Will this result
 % in a false positive or a false negative or what?}

 % RB: This means that the algorithms do not appear
 % in the CrySL spec, which will generate a warning.
 % Don't think it is a false positive or false
 % negative.
 
 We also found some differences when considering the
 particular platform versions. For instance, until version
 10 of the Android platform, developers must use the
 \texttt{TLS}\footnote{TLS is a protocol that provides
authenticated encryption for data connections.} algorithm
 for \texttt{SSLContext}. This led to 169 additional violations regarding the
 incorrect usage of the \texttt{SSLContext} class in the
 ``Android Base 0108'' configuration, in comparison
 to ``Android Base 0116'' and ``Android Base 25plus''.
 In more detail, 161 apps use either the \texttt{SSL} or
 \texttt{TLSv1} algorithms (both introduced
 in version \texttt{10}) and eight apps use either
 \texttt{TLSv1.1} or \texttt{TLSv1.2} (both introduced
 in version \texttt{10}). These violations do not
 occur in the remaining ``0116'' and ``25plus'' configurations.
 We also found similar divergences on the platform
 version related to other cryptographic primitives.

 It is important to note that, although
 version 8 was released in May 2010 already, in order to
 increase compatibility with a broader range of devices, most 
 apps in our dataset are still configured to use this version as the minimum version. 
 The observation that the number of violations for the ``Android Base 0108'' configuration is higher compared to the
 the ``Android Base 0116'' and ``25plus''
 configurations might indicate that some apps use cryptographic algorithms that are not available in their minimum version. This would then lead to a runtime exception.
In summary: 

 \begin{mh}
The experiments showed a
   significant difference when considering the different
   versions of the platform for the Android Base configurations.
   Yet, the Android Base configurations are much less restrictive then those of the BSI and by CogniCrypt in general.
We found slight differences in the results of
   \ca when considering the recommendations
   from BSI and CogniCrypt. Although the differences
   are not that large, this result still suggests that
   one can benefit from tailoring the specifications of the
   correct usage of cryptographic standards 
   according to the different guidelines. 
\end{mh}

\section{Threats to Validity}

In this section, we present some limitations and
possible threats to the validity of our work. 
Since our research focuses on cryptographic libraries only, we need to discuss the applicability of our approach to other domains.
The choice of this domain was motivated by our previous experience using
\csl to specify the correct usage of cryptographic APIs. We was challenged 
by the fact that new algorithms
are frequently designed and old ones might
become deprecated~\cite{cryptoevol}. In addition, cryptographic
standards are frequently updated---in particular to state that
an algorithm vulnerability has been found and reported.

We believe that our approach can
also be used for APIs that target other domains as well, even though
we did not systematically investigate this
question.
First, APIs from different domains evolve along the time,
and as we discussed throughout this paper, API evolution has an impact
on the correct usage of libraries.
%% API evolution is a
%% recurrent challenge, and
%% there is a large body of knowledge exploring
%% this issue (e.g.,~\cite{chow:icsm-96,mezini:ecoop1997,schafer:icse-2008,robes:fse-2012}).
Second, there are different recommendations
on the proper usage of each popular API. For instance,
there are many guidelines discussing the correct usage of the Java
Persistence API~\cite{durstin:jpa,leonard2020spring}---and individual companies might
also take advantage of specific recommendations. The
specifications about how to correctly use a given
API should take into account these differences. 
We envision that both practitioners and researchers
would benefit from a domain engineering approach
that considers different sources of variability---
including different versions of an API, recommendations from gray
literature (for instance), and mining software
repositories efforts---before specifying the correct usage
a given API. We make the reader aware that domain engineering is a well-known technique to understand properties that,
like in our case, bring variability to the domain of API usage specifications.
We are not attempting to validate  domain engineering itself or propose a technique for its application to other domains;
the process for which would require careful understanding of the specific API domain and a
thorough analysis. 

%% RB: I am commenting the paragraph bellow because
%% we have omitted the MSR study on cryptography
%% API usage. 

%% \eb{Again this does not describe a threat to validity. Do you mean to say that the external validity of the experiment is
%% threatened by the fact that we only considered the JCA and then justify that choice?}

%% Although there are several cryptographic libraries for
%% Java, only the JCA presents a widespread
%% usage in open source Android and Java projects. Curiously,
%% although Google Tink is a popular project (in terms
%% of number of GitHub stars), popular Java and Android
%% projects do not use this library. One possible reason for
%% not finding projects using Google Tink is that it
%% is a recent library, with a first public
%% release (Tink 1.0.0) available in 2017. Another possible explanation
%% is that here we only focus on ``popular'' and ``large''
%% GitHub projects---and this might also affect the
%% external validity of our research. Therefore, we could not
%% generalize the results of our MSR phase (Section~\ref{sec:msr}) to
%% projects that do not fit in our exclusion criteria
%% (size and popularity) or hosted in other platforms
%% (e.g., SourceForge, GitLab, etc.).

Another threat to our conclusions relates to the additional complexity introduced by \mcsl. We envision that the users of \mcsl are already users of CrySL, and the learning curve would involve a language for specifying \csl refinements and configurations. To better quantify the additional complexity \mcsl introduces, we will have to conduct a \emph{user study} with this specific goal. We postpone such an investigation to a future work, since our focus here was to explore \mcsl in a more realistic scenario, investigating the possible benefits of using \mcsl to modularize \csl specifications for different versions of the Android platform and different cryptographic recommendations. Therefore, currently we do not have empirical evidence about how much complexity \mcsl introduces to those already familiar with \csl. Nonetheless, compared to the benefit of managing a large family of specifications using a relatively small number of refinements and configurations, we feel this additional complexity is justified.

Additional threats relate to the
methods we used in our research. We
tried to mitigate possible \emph{reliability threats}
by reusing methods and tools from previous research studies.
For instance, we investigated the frequency of \emph{breaking
changes} to estimate the stability of Java
cryptographic libraries using the methodologies
available in the literature~\cite{api-diff,brito:saner2018,xavier:saner2017}.
Nonetheless, although we found more than 1700 \emph{breaking changes} across 11
public releases of Bouncy Castle, a limitation of our work 
is that we do not investigate if these changes have
actually broken existing client code. Our understanding
is that just a subset of breaking changes impact on
the specifications of the correct usage of APIs.

This threat relates to the use of \apidiff, which
detects breaking changes considering modifications
to the \emph{standard notion} of Java interfaces---that is,
public members of Java classes or interfaces. Modifications
that do not preserve the standard notion of Java interfaces
(e.g., changing the signature of public methods, removing public methods, and so on)
are claimed by \apidiff as breaking changes. This
might actually lead to a number of false-positives---once
client code might not depend on all public members of
a library. To mitigate this threat, we narrowed our
analysis of the Bouncy Castle library to only focus 
on the high-level classes and interfaces of Bouncy Castle
that implement cryptographic primitives.

Regarding our research question \textbf{RQ4}, we measure the reduction of lines of
specification and redundancy
with respect to \emph{generated} specifications.
This might raise the question whether these generated specifications do not contain boilerplate text that had not arisen had these
specifications be hand-written. We are confident that we can rule this out, due to the nature of \csl specifications
and the way they are generated by \mcsl. Conducting a large scale developer study by manually writing many families of specifications by hand was beyond the scope of this work. 

%% In order to
%% allow other researchers to reproduce our study, we
%% also  make available a replication package with
%% the datasets, scripts, and tools used in this research. 

%\sk{Internal:}
%\sk{The selection of \csl configurations in Table 4.}
%\sk{We didn't manually verify \ca's results (manually or othwerwise).}

\section{Related Work}

\subsection{Domain Engineering}

Frakes et al.~\cite{frakes1998dare} present a well-established
definition for domain engineering, which
embraces two phases: \emph{domain analysis} and
\emph{domain implementation}. The first deals
with all activities necessary to understand
and document the commonalities and variabilities
within a software domain. Similar to
the guidelines presented by the authors, we also collected
and recorded information from documents (cryptographic
standards) and source code (examples of cryprographic
libraries usage) while conducting our domain analysis.
The main difference of our approach is that we
also mined the source code
evolution of the cryptographic libraries.
Lisboa et al. presents a literature review on
tools and methods for domain analysis~\cite{lisboa:ist-2010}. 

The second phase of domain engineering (that is,
domain implementation) aims to build
the infrastructure necessary
to generate products from reusable assets.
Here we used the same general idea, though not to
build software products, but actually to
generate specifications of the correct usage
of APIs that might vary according to
different sources of variablity (such as
versions of APIs, platforms, and cryptographic
standards). Czarnecki and Eisenecker~\cite{gp-book} detail
several techniques that can be used to
implement an infrastructure for building
products from reusable assets. In our work, we
used the \emph{refinement-based} transformational
approach~\cite[Chapter~9]{gp-book} as the basis for the \mcsl design
and implementation. The literature on software
product lines also recommends two distinct
phases for building SPLs: one for domain analysis and
one for domain implementation~\cite{apel-spl-book,spl-book}.

\subsection{Correct Usage of APIs}

Amann et al. present some terminology and
taxonomy around the correct usage and
misuse of APIs~\cite{amann:tse-2019}.
Given a set of constraints stating,
for instance, the expected order of method
calls and the pre-conditions the client
code must guarantee before calling the
methods of an API, any usage scenario
that violates a constraint characterizes
a misuse---otherwise, it is a correct usage~\cite{amann:tse-2019}.
The main goal of mining misuse of APIs is
to reveal \emph{deviant code} that might
originate a bug or a software vulnerability
(in the context of cryptographic APIs).

According to Amann et al~\cite{amann:tse-2019},
the constraint specifications could be
manually crafted by experts or infered
using either dynamic~\cite{pradel:icse-2012,luo:rv-2014} or static
analysis~\cite{wasylkowski:fse-07,monperrus:tosem-2013,saied:saner-2015}.
In this paper, we rely on a manually crafted approach
to specify rules in \mcsl---mostly because
many programs fail to use cryptographic APIs
correctly~\cite{nadi:icse2016,acar:sp2017,kruger:tse2019}.
It is a matter of future work to investigate
if our domain engineering approach could also benefit from
techniques that automatically infer the correct
usage of APIs. 

To the best of our knowledge, none of
the previous research works consider that the
correct usage of an API could vary, among
other reasons, according
to specific versions of APIs or to existing
usage recommendation patterns that could be general
accepted or tailored to particular companies or
projects.

\subsection{API Evolution}

Studies on API evolution focus on two directions.
First, to help developers to migrate their
systems in response to the evolution
of APIs the systems depend
on~\cite{chow:icsm-96,mezini:ecoop1997,schafer:icse-2008,henkel:icse-2005}.
The second direction, which
is closely related to our research, focus on understanding
how developers evolve APIs and on characterizing
the evolution of APIs. For instance, several
research works have explored the impact of
\emph{deprecation} mechanisms on software
ecosystems~\cite{robes:fse-2012,swant:icsme-2016,sawant2018reaction}.
Other research studies investigate how developers respond
to API evolution~\cite{hora:icsme-2015} and the motivations
for breaking APIs~\cite{brito:ese-2020}.

Here we investigate how the evolution of
cryptographic APIs occurs in practice,
considering the history of three Java
cryptographic libraries: JCA/JCE, Bouncy
Castle, and Google Tink. We have found
that cryptographic libraries are quite
stable, and the high-level APIs that
define cryptographic primitives rarely
change---even though we found a number
of \emph{breaking changes} during the
evolution of Bouncy Castle. The most
typical pattern is the introduction of
new algorithms that implement cryptographic
primitives---which often requires changes
into the specification about the correct
usage of the APIs. 

\section{Conclusion}

Domain engineering involves a set
of techniques for identifying and
documenting the commonalities and
variabilities within a software domain,
as well as for building an infrastructure
for deriving products from reusable assets~\cite{apel-spl-book,frakes1998dare,spl-book}.
While it has been successfully used to develop software product
lines, in this paper, we explored the use of
domain engineering procedures to
specify the correct usage of cryptographic
APIs. After gathering a better understanding
about how different versions of the
platforms, APIs, and cryptographic
standards might affect the specifications
of the correct usages of crypto APIs, we designed
\mcsl. \mcsl serves as an infrastructure for generating \csl~\cite{kruger:tse2019}
specifications tailored for specific
scenarios. We evaluated our approach using
a family of \mcsl specifications describing the correct
usage of the Java Cryptographic Architecture for
Android, which accommodates the evolution of
the Android platform and three distinct
sets of cryptographic recommendations.
Our results provide evidence that
it is important to tackle the problem of
writing specifications of correct usage of
APIs using a domain engineering approach and
that using \mcsl we can better modularize
families of specifications.

%
% The acknowledgments section is defined using the "acks" environment (and NOT an unnumbered section). This ensures
% the proper identification of the section in the article metadata, and the consistent spelling of the heading.
%\begin{acks}
% We should not forget to put the CROSSING acks here.
%\end{acks}

%
% The next two lines define the bibliography style to be used, and the bibliography file.
% \bibliographystyle{ACM-Reference-Format}
\bibliography{references}

% 
% If your work has an appendix, this is the place to put it.
\appendix

\end{document}